\documentclass{article}
\usepackage{amssymb,graphicx}

\def\prb{Phys.\ Rev.\ B\ }
\def\pra{Phys.\ Rev.\ A\ }
\def\prl{Phys.\ Rev.\ Lett.\ }
\def\rmp{Rev.\ Mod.\ Phys.\ }
 
\begin{document}

\title{Stripe and bubble phases in quantum Hall systems%
\footnote{To be published in
{\it High Magnetic Fields: Applications in Condensed Matter Physics
and Spectroscopy\/} (Springer-Verlag, Berlin, 2002).}
}
\author{Michael M. Fogler}
\maketitle

\noindent{Department of Physics, Massachusetts Institute of
Technology, 77 Massachusetts Avenue, Cambridge, MA 02139, USA}
\\[6pt]

\begin{abstract}

We present a brief survey of the charge density wave phases of a
two-dimensional electron liquid in moderate to weak magnetic fields
where several higher Landau levels are occupied. The review follows the
chronological development of this new and emerging field: from the ideas
that led to the original theoretical prediction of the novel ground
states, to their dramatic experimental discovery, to the currently
pursued directions and open questions.

\end{abstract}

\section{Historical background}
\label{Background}

Until recently, the quantum Hall effect research effort has been focused
on the case of very high magnetic fields where electrons occupy only the
lowest and perhaps, also the first excited Landau levels (LL).
Investigation of the weak magnetic field regime where higher LLs are
populated, was not considered a pressing matter because no particularly
interesting quantum features could be discerned in the magnetotransport
data. The experimental situation has changed around 1992, when extremely
high purity two-dimensional (2D) electron systems became available. At
low temperatures, $T < 30 {\rm m K}$, such samples would routinely
demonstrate very deep resistance minima at integral filling fractions
down to magnetic fields of the order of a tenth of a
Tesla~\cite{Coleridge_94}. This indicated that the quantum Hall effect
could persist up to very large LL indices, such as $N \sim 100$, and
called for the theoretical treatment of the high LL problem. Although
the integral quantum Hall effect could be explained without invoking
electron-electron interaction, two other experimental findings strongly
suggested that the interaction {\it is\/} important at large $N$. One
was the prominent enhancement of the bare electron
$g$-factor~\cite{Rokhinson} and the other was a pseudogap in the tunneling
density of states~\cite{Turner_96}. Surprisingly, {\it no\/}
interaction-induced fractional quantum Hall effect has ever been
observed at higher $N$ in contrast to the case of the lowest and the
first excited LLs ($N = 0$ and $1$). An effort to understand this
puzzling set of facts led A.~A.~Koulakov, B.~I.~Shklovskii, and the
present author to the theory of charge density wave phases in partially
filled $N \ge 2$ LLs~\cite{Fogler_96,Fogler_97}. When this theory
received a dramatic experimental support~\cite{Lilly_99,Du_99}, a broad
interest to the high LL physics has emerged. Below we give a brief
review of this new and exciting field. Some of the ideas presented here
are published for the first time. For previous short reviews on the
subject see Refs.~\cite{Eisenstein_01,Oppen_00}.

\section{Landau quantization in weak magnetic fields}
\label{Quantization}

\begin{figure}
\includegraphics[width=4.5in,bb=114 336 488 530]{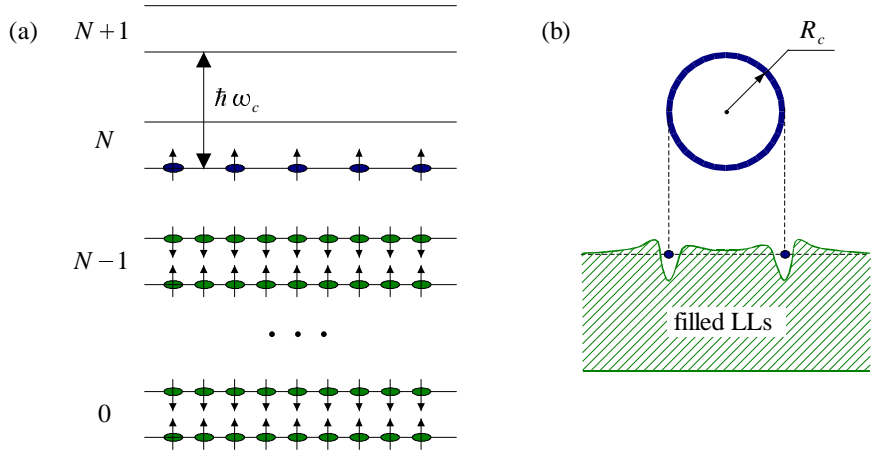}
\vspace{0.1in}
\caption{
(a) Landau levels. Darkened ellipses symbolize electrons and arrows ---
their spins (b) A quasiparticle at the $N$th LL viewed as a ring-shaped
object immersed into a medium formed by the filled lower LLs.
\label{Fig_rings}
}
\end{figure}

Consider a 2D electron system with the areal density $n$ in the presence
of a transverse magnetic field $B$. For the case of Coulomb interaction,
at zero temperature and without disorder, the properties of such a
system are determined by exactly three dimensionless parameters: $r_s$,
$\nu$, and $E_Z / \hbar \omega_c$. The first of these, $r_s = (\pi n
a_B^2)^{-1/2}$, measures the average particle distance in units of the
effective Bohr radius $a_B = \hbar^2 \kappa / m e^2$. The properties of
the electron gas are very different at large and small $r_s$. In these
notes will focus exclusively on the case $r_s \lesssim 10$. This is
roughly the condition under which the electron gas in zero magnetic
field behaves as a Fermi-liquid~\cite{Mahan_book}. As we discuss below,
the system is no longer a Fermi-liquid at any finite $B$; however, the
basic structure of LLs separated by the gaps $\hbar \omega_c$, where
$\omega_c = e B / m c$ is the cyclotron frequency, survives at arbitrary
low $B$. In this situation, the second dimensionless parameter, $\nu = 2
\pi l^2 n$, specifies how many LL subbands are occupied. Here $l =
(\hbar c / e B)^{1/2}$ is the magnetic length. The lower LLs are fully
occupied, while the topmost level is, in general, partially filled.
Therefore, $\nu = 2 N + \nu_N$, where the factor of two accounts for the
spin degree of freedom and $\nu_N$ is the filling fraction of the
topmost ($N$th) LL, $0 < \nu_N < 2$ (see a cartoon in
Fig.~\ref{Fig_rings}a).

The remaining dimensionless parameter $E_Z / \hbar \omega_c$ introduced
above is the ratio of the Zeeman and the cyclotron energy. It affects
primarily the dynamics of the spin degree of freedom, which is beyond
the scope of these notes. Suffices to say that in the ground state the
topmost $N$th LL is thought to be fully spin-polarized for $N >
0$~\cite{Wu_95} with a sizeable spin gap. In the important case of GaAs,
the spin gap greatly exceeds $E_Z$ due to many-body effects.

Because of the spin and the cyclotron gaps, the low-energy physics is
dominated by the electrons residing in the single spin subband of a
single (topmost) LL. All the other electrons play the role of an
effective dielectric medium, which merely renormalizes the interaction
among the ``active'' electrons of the $N$th LL. This elegant physical
picture was first put forward in an explicit form by Aleiner and
Glazman~\cite{Aleiner_95}.

The validity of such a picture in weak magnetic fields is
certainly not obvious. Naively, it seems that as $B$ and $\hbar\omega_c$
decrease, the LL structure should eventually be washed out by the
electron-electron interaction. The following reasoning shows that this
does not occur (see also Ref.~\cite{Aleiner_95} for somewhat different
arguments). Let us divide all the interactions into three groups: (a)
intra-LL interaction within $N$th level, (b) interaction between the
electrons of $N$th level and its near neighbor LLs, and (c) interaction
between $N$th and remote LLs (with indices $N^\prime$ significantly
different from $N$). The last group of interactions is characterized by
frequencies much larger than $\omega_c$. It is not sensitive to the
presence of the magnetic field and leads only to Fermi-liquid
renormalizations of the quasiparticle properties. Interactions within
the groups (a) and (b) have roughly the same matrix elements but the
latter are suppressed because of the cyclotron gap. Hence, it is the
interactions among its own quasiparticles that are the most
``dangerous'' for the existence of a well-defined $N$th LL. It is crucial
that the intra-LL interaction energy scale does not exceed the typical value
of
\begin{equation}
                  E_{\rm ex} \sim 0.1 e^2 / \kappa R_c,
\label{E_int}
\end{equation}
where $R_c = \sqrt{2 N + 1}\, l$ is the classical cyclotron
radius~\cite{Aleiner_95,Fogler_96}. The ratio $E_{\rm ex} /
\hbar\omega_c \sim 0.1 r_s$ is $B$-independent; thus, there is a good
reason to think that the validity domain of the proposed
single-Landau-level approximation extends down to arbitrary small $B$'s
and, in fact, is roughly the same as that of the Fermi-liquid ($r_s
\lesssim 10$). This is certainly borne out by all available
magnetoresistance data~\cite{Coleridge_94,Rokhinson,Lilly_99,Du_99}.
Henceforth we focus exclusively on the quasiparticles residing at the
topmost LL.

The inequality $E_{\rm ex} < \hbar \omega_c$ means that the cyclotron
motion is the fastest motion in the problem, and so on the timescale at
which the ground-state correlations are established, quasiparticles
behave as clouds of charge smeared along their respective cyclotron
orbits, see Fig.~\ref{Fig_rings}b. The only low-energy degrees of
freedom are associated with the guiding centers of such orbits. In the
ground state they must be correlated in such a way that the interaction
energy is the lowest. This prompts a quasiclassical analogy between the
partially filled LL and a gas of interacting ``rings'' with radius $R_c$
and the areal density $\nu_N / (2 \pi l^2)$. Note that for $\nu_N > 1 /
N$ the rings overlap strongly in the real space.

Strictly speaking, the guiding center can not be localized a single
point, and so our analogy is not precise. However, the quantum
uncertainty in its position is of the order of $l$. At large $N$, where
$l \ll R_c$, the proposed analogy becomes accurate and useful. For
example, it immediately clarifies the physical meaning of $E_{\rm ex}$
as a characteristic interaction energy of two overlapping rings.

%
%
\setlength{\columnwidth}{4in}

Technically, the high LL problem is equivalent to the more
studied $N = 0$ case if the bare Coulomb interaction $\tilde{v}_0(q)$ is
replaced by the renormalized interaction
\begin{equation}
                  \tilde{v}(q) = \frac{\tilde{v}_0(q)}{\epsilon(q)}
\left[L_N\left(\frac{q^2 l^2}{2}\right)\right]^2,
\label{v_q}
\end{equation}
where $\epsilon(q)$ is the dielectric constant due to the screening by
other LLs~\cite{Kukushkin_88,Aleiner_95} and the bracketed expression
compensates for the difference in the form-factor
\begin{equation}
                  F_N(q) = L_N\left(\frac{q^2 l^2}{2}\right)
                           e^{-q^2 l^2 / 4}
\label{F_N}
\end{equation}
of the cyclotron orbit at $N$th and at the lowest LLs, with
$L_N(z)$ being the Laguerre polynomial. In the next section we will
discuss the consequences of having such an unusual interaction.

\section{Charge density wave instability}
\label{Instability}

The mean-field treatment of a partially filled LL amounts to the
Hartree-Fock approximation, first examined in the present context by
Fukuyama~{\it et al\/}. It is worth pointing out the differences between
their paper~\cite{Fukuyama_79} and our own work~\cite{Fogler_96}
reviewed below in this section. The pioneering work of Fukuyama {\it et
al.\/}~\cite{Fukuyama_79} appeared in 1979. A few years later, after the
discoveries of integral and fractional quantum Hall effects, it became
clear that for the exception of dilute limit, the Hartree-Fock
approximation is manifestly incorrect for the lowest LL
case~\cite{FQHE}. By 1995 when we started to work on the high LL
problem, Ref.~\cite{Fukuyama_79} has been effectively shelved away. In
contrast to Ref.~\cite{Fukuyama_79}, who did not try to assess the
validity of the Hartree-Fock approximation, our theory of a partially
filled high LL~\cite{Fogler_96} was based on this kind of approximation
because it {\it is\/} the correct tool for the job. This point is
elaborated further in Sec.~\ref{HF_validity}. Another important
difference from Ref.~\cite{Fukuyama_79} is a parametric dependence of
the wavevector $q_*$ of the CDW instability on the magnetic field: we
find $q_* \propto B^{-1}$ instead of $\propto B^{-1/2}$ in the theory of
Fukuyama~{\it et al\/}. Finally, the physical picture of ring-shaped
quasiparticles that guided our intuition is quite novel and applies only
for high LLs.

\begin{figure}
\includegraphics[width=3.0in,bb=143 495 496 709]{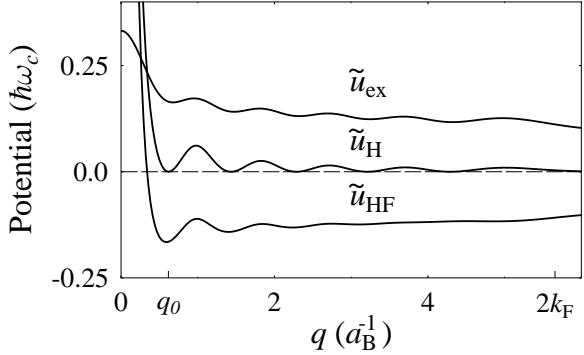}
\caption{
Direct, exchange, and the total Hartree-Fock potentials in
$q$-space for $N = 5$ and $r_s = 0.5$. [Reproduced with permission
from Fig.~2 of Ref.~\cite{Fogler_96} (a)].
\label{Fig_u}
}
\end{figure}

Within the Hartree-Fock approximation, the free energy of the system is
given by
\begin{equation}
   F_{\rm HF} = \frac{1}{4 \pi l^2} \sum_{q}
                \tilde{u}_{\rm HF}(q)
                |\langle \tilde\Delta(q) \rangle|^2
              + k_B T \sum_X \langle n_X \ln n_X
              + (1 - n_X) \ln(1 - n_X) \rangle,
\label{F_HF}
\end{equation}
where $\tilde\Delta(q) = (2 \pi l^2 / L_x L_y) \sum_X e^{- i q_x X}
a^\dagger_{X + q_y l^2 / 2} a_{X - q_y l^2 / 2}$ are guiding center
density operators, $L_x$ and $L_y$ are system dimensions, $a^\dagger_X$
($a_X$) are creation (annihilation) operators of Landau basis states $|
X \rangle$, and $n_X = a^\dagger_X a_X$ are their occupation numbers
subject to the constraint $\sum_X \langle n_X \rangle = L_x L_y \nu_N /
(2 \pi l^2)$. Here and below we assume that $0 < \nu_N < 1$ because the
states with $1 < \nu_N < 2$ are the particle-hole
transforms of the states with $2 - \nu_N$ and do not require a
special consideration. The Hartree-Fock interaction potential is defined
by $\tilde{u}_{\rm HF}(q) = (1 - \delta_{q, 0}) \tilde{u}_H(q) -
\tilde{u}_F(q)$, where
\begin{equation}
   \tilde{u}_H(q) = \tilde{v}(q) e^{-q^2 l^2 / 2},\quad
   \tilde{u}_{\rm ex}(q) = 2 \pi l^2 u_H(q l^2) 
 \label{u_HF}
\end{equation}
are its direct and exchange components (tildes denote Fourier
transforms)~\cite{Fogler_96}. Their $q$-dependence is illustrated in
Fig.~\ref{Fig_u}.

%
%
\setlength{\columnwidth}{4in}

The inherent feature of $\tilde{u}_{HF}(q)$ is a global minimum at
\begin{equation}
q_* \approx 2.4 / R_c,
\label{q_optimal}
\end{equation}
where it is negative, $\tilde{u}_{\rm HF}(q_*) < 0$. Within the
Hartree-Fock theory, it leads to a charge density wave (CDW) formation
at low enough temperatures $T$~\cite{Fukuyama_79}. Indeed, at high
temperatures the entropic term dominates and the equilibrium state is a
uniform uncorrelated liquid with $\langle n_X \rangle = \nu_N$ and
$\langle \tilde\Delta(q) \rangle = \delta_{q, 0} \nu_N$. At lower $T$,
it is more advantageous to forfeit some entropy but gain some
interaction energy by creating a guiding center density modulation
$\langle \tilde\Delta_q \rangle \neq 0$ with wavevector $q = q_*$.

Since the quasiparticles are extended ring-shaped objects, the actual
quasiparticle density modulation in any of our states is given by the
product of the amplitude $\tilde\Delta(q)$ of the guiding center density
wave and the cyclotron orbit form-factor:
\begin{equation}
                    \tilde\rho(q)  = \tilde\Delta(q) F_N(q).
\label{rho_and_Delta}
\end{equation}
The physical electric charge modulation in the system is further
suppressed by the additional factor of $\epsilon(q)$ due to the
screening by the lower LLs. A peculiarity of the $N \gg 1$ case is that
$q_*$ is very close to the first zero $q_0$ of $\tilde{u}_H(q)$, which
it inherits from the form-factor $F_N(q)$.~\footnote{In contrast, in the
$N = 0$ case studied by Fukuyama~{\it et al.\/}, $\tilde{u}_{\rm H}(q)$
does not have nodes.} On the one hand, this means that the physical
electric charge modulation is always rather small --- a few percent in
realistic experimental conditions. On the other hand, it explains why
this instability develops in the first place. Indeed, usually the direct
electrostatic interaction is repulsive and dominates over a weak
attraction due to exchange. In our system $\tilde{u}_H(q)$ vanishes at
the ``magic'' wavevector $q_0$ because no charge density is induced,
$\rho(q_0) = 0$. As a result, the exchange part dominates and gives rise
to a range of $q$'s around $q_0$ where the net effective interaction
$\tilde{u}_{\rm HF}(q)$ is attractive, which leads to the instability.


The nodes of $F_N(q)$ responsible for the vanishing of $\rho(q)$ exist
for a purely geometric reason that the quasiparticle orbitals are
extended objects of a specific ring-like shape. The size of the orbitals
is uniquely defined by the total density and the total filling factor.
These two facts make the position of the global minimum $q_*$ very
insensitive to approximations contained in the Hartree-Fock approach as
well as many microscopic details, e.g., the functional form of
$\epsilon(q)$, thickness of the 2D layer, which affects
$\tilde{v}_0(q)$, {\it etc.} \/At asymptotically large $N$ where the
rings are very narrow, $F_N(q)$ is closely approximated by a Bessel
function $J_0(q R_c)$; hence, Eq.~(\ref{q_optimal}). Surprisingly,
Eq.~(\ref{q_optimal}) is quite accurate even at $N \sim 1$.

\begin{figure}
\includegraphics[width=2.0in,bb=225 393 379 505]{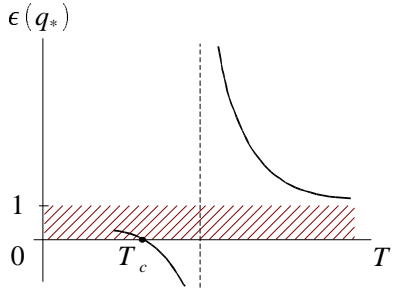}
\caption{
$T$-dependence of the dielectric function of the liquid
state. The unstable region is hatched.
\label{Fig_epsilon_tot}
}
\end{figure}

The mean-field transition temperature $T_c^{\rm mf}$ can be estimated as
the point where the stability criterion $1 / \epsilon_{\rm tot}(q) < 1$
of the uniform liquid state is first violated. Here $\epsilon_{\rm
tot}(q)$ is the total dielectric function, including both the lower and
the topmost LLs. A simple derivation~\cite{Fogler_unpub} within the
(time-dependent) Hartree-Fock approximation gives
\begin{equation}
          \epsilon_{\rm tot}(q) = \epsilon(q)
\left\{
1 + \tilde{u}_H(q) \left[\frac{2 \pi l^2 k_B T}{\nu_N (1 - \nu_N)}
- \tilde{u}_{\rm ex}(q)\right]^{-1} \right\}.
\label{epsilon_tot}
\end{equation}
The $T$-dependence of this function at $q = q_*$ is sketched in
Fig.~\ref{Fig_epsilon_tot}.

%
%
\setlength{\columnwidth}{4in}

The stability criterion leads to the estimate
\begin{equation}
          k_B T_c^{\rm mf} = \frac{\nu_N (1 - \nu_N)}{2 \pi l^2}
                         |\tilde{u}_{HF}(q_*)|.
\label{T_c_mf}
\end{equation}
For $r_s \sim 1$, $N \gg 1$, and $\nu_N = 1 / 2$, it can be
approximated by
\begin{equation}
 k_B T_c^{\rm mf}(\nu_N = 1 / 2) \approx 0.02 \hbar\omega_c.
\label{T_c}
\end{equation}
Compared to the expression originally given in Ref.~\cite{Fogler_96}(b),
a certain $1 / N$ term is omitted here, out of precaution that the
Hartree-Fock approximation, which is valid in the large-$N$ limit, may
not have enough accuracy to describe $1 / N$-corrections. This caveat
should always be kept in mind when using numerical estimates of $k_B
T_c^{\rm mf}$, such as those reported in Ref.~\cite{Stanescu_00}. For
the typical experimental situation~\cite{Lilly_99,Du_99},
Eq.~(\ref{T_c_mf}) gives $T_c^{\rm mf} \sim 1 {\rm K}$. 

\section{Mean-field phase diagram}
\label{Mean_field_phases}

\subsection{Charge density wave transition}

A more systematic way to study the CDW transition is via a Landau
expansion of the free energy $F$ in powers of the order
parameters $\langle \tilde\Delta({\bf q}) \rangle$ where ${\bf q}$ are
restricted to the locus of the soft modes, $|{\bf q}| = q_*$:
\begin{equation}
 F = \sum\limits_{n = 2}^\infty a_n(T)
\sum\limits_{{\bf q}_1 + {\bf q}_2 + \ldots + {\bf q}_n = 0}
\prod\limits_{i = 1}^{n} \langle \tilde\Delta({\bf q}_i) \rangle.
\label{Landau_expansion}
\end{equation}
Note that in the quasiclassical large-$N$ limit, the order parameter
$\langle \Delta({\bf r}) \rangle$ is proportional to the local filling
factor, $\langle \Delta({\bf r}) \rangle = \nu_N({\bf r}) / 2 \pi l^2$.

\begin{figure}
\includegraphics[width=4.0in,bb=153 409 449 564]{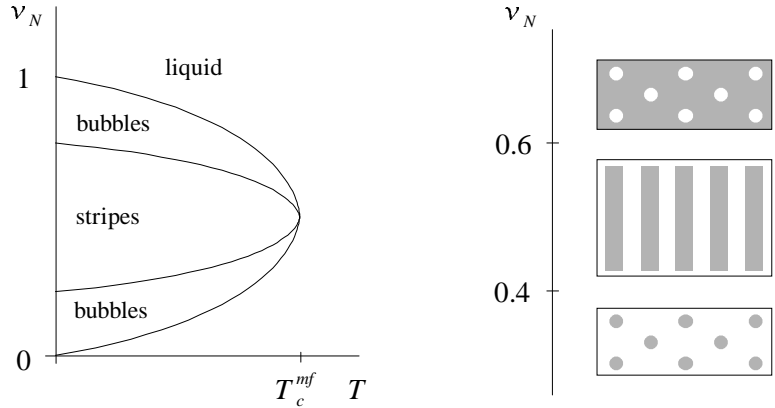}
\vspace{0.1in}
\caption{
Left: Mean-field phase diagram.
Right: Guiding center density domain patterns at $T = 0$. Shaded and
blank areas symbolize filled and empty regions, respectively.
\label{Fig_stripes_and_bubbles}
}
\end{figure}

The above linear stability criterion is equivalent to the condition $a_2
> 0$. At $\nu_N = \frac12$ where the cubic term vanishes by symmetry,
$a_3 = 0$, the Landau theory's estimate for $T_c$ coincides with
Eq.~(\ref{T_c_mf}). It predicts a second-order
transition~\cite{Fukuyama_79,Moessner_96}, which occurs by a
condensation of a single pair of harmonics, whose direction is chosen
spontaneously, e.g., ${\bf q} = \pm q_* \hat{\bf x}$. The resultant
low-temperature state is a unidirectional CDW or the {\it stripe phase}.
Away from the half-filling, $a_3 \neq 0$. Hence, at $\nu_N \neq \frac12$
the transition is of the first order, takes place at a temperature
somewhat higher than predicted by Eq.~(\ref{T_c_mf}), and is from the
uniform liquid into a CDW phase with the triangular lattice
symmetry~\cite{Fukuyama_79,Moessner_96}, the {\it bubble phase\/}, see
Fig.~\ref{Fig_stripes_and_bubbles} (left).

%
%

Near its onset, the CDW order brings about only a small modulation of
the local filling factor, so that the topmost LL remains partially
filled everywhere. As $T$ decreases, the amplitude of the guiding center
density modulation increases and eventually forces expulsion of regions
with partial LL occupation. The system becomes divided into (i)
depletion regions where $\langle \Delta({\bf r}) \rangle = 0$ and local
filling fraction is equal to $2 N$, and (ii) fully occupied areas where
$\langle \Delta({\bf r}) \rangle = (2 \pi l^2)^{-1}$ and local filling
fraction is equal to $2 N + 1$ (however, see Ref.~\cite{Fogler_96} for a
discussion of truly small $r_s$). At these low temperatures the {\it
bona fide\/} stripe and bubble domain shapes become evident, see
Fig.~\ref{Fig_stripes_and_bubbles} (right).

\subsection{Stripe to bubble transition}

\begin{figure}
\includegraphics[width=4.2in,bb=44 425 530 631]{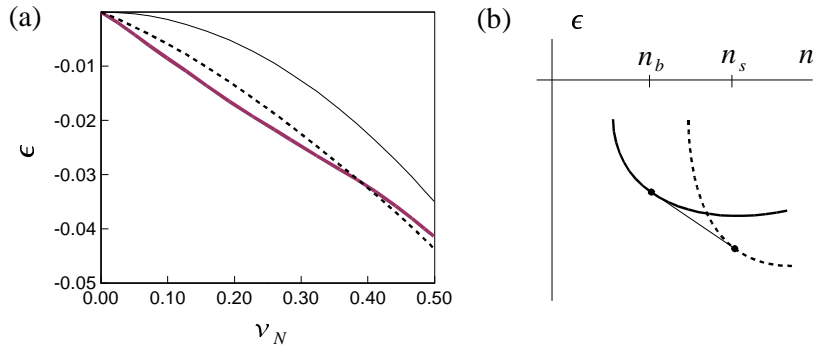}
\caption{
(a) The energy density $\epsilon$ in units of $\hbar\omega_c / 2 \pi
l^2$ as a function of $\nu_N$ for the bubbles (thick solid line), the
stripes (dashed line), and the uniform liquid (thin solid line) for $r_s
= 1$, $N \gg 1$, and $T = 0$. (b) Schematics of the conventional tangent
construction. As explained in the main text, it is too crude to account
for the specifics of the long-range Coulomb interaction. For example,
for the graph on the left, such a construction would give a vanishing
$n_b$ and the maximum possible $n_s$, which is misleading.
\label{Fig_stripe_and_bubble_energy}
}
\end{figure}

Near $T_c$, the Landau theory~\cite{Moessner_96} predicts the
stripe-bubble transition to be of the first order. This seems to be the
case at $T = 0$ as well, at least at large $N$, where the this
transition occurs at $\nu_N \approx 0.39$~\cite{Fogler_96}, see
Fig.~\ref{Fig_stripe_and_bubble_energy}a. In systems with only
short-range interactions a density-driven first order transition is
accompanied by a global phase separation. An example is the usual
gas-liquid transition. The densities $n_s$ and $n_b$ of the two
co-existing phases are determined by Maxwell's tangent construction,
Fig.~\ref{Fig_stripe_and_bubble_energy}b.

%
%

\begin{figure}
\includegraphics[width=2.1in,bb=224 434 374 546]{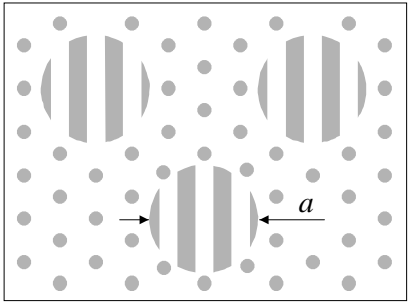}
\caption{
A cartoon of the conjectured superbubble phase.
\label{Fig_super}
}
\end{figure}

In the present case the long-range Coulomb interaction changes the
situation drastically. The macroscopic phase separation into two phases
of different charge density is forbidden by an enormous Coulomb energy
penalty. Only a phase separation on a finite lengthscale, i.e., domain
formation may occur. Since stripes and bubbles are two the most common
domain shapes in nature~\cite{Seul_95}, it opens an intriguing
possibility that bubble- or stripe-shaped domains of the stripe phase
inside of the bubble phase may appear, i.e., the ``superbubbles''
(Fig.~\ref{Fig_super}) or the ``superstripes''. Their size would be
determined by the competition between the Coulomb energy and the domain
wall tension $\gamma$. The superbubbles would have a diameter
\begin{equation}
 a \sim \frac{1}{n_s - n_b} \left(\frac{\gamma \kappa}{e^2}\right)^{1/2}
\label{a}
\end{equation}
and increase the net energy density by $\sim \gamma p / a$, $p$ being
the fraction of the minority phase. It turns out that this additional
energy cost shrinks the range of the phase co-existence considerably
compared to that in the conventional tangent construction. It may
totally preclude the phase co-existence if
\begin{equation}
 \frac{e^2 \gamma}{\kappa} > (n_s - n_b)^2
              \frac{d^2 f_s}{d n^2} \frac{d^2 f_b}{d n^2},
\label{phase_coexistence}
\end{equation}
where $f_{s(b)}(n)$ is the free energy density of stripes (bubbles).

Incidentally, these considerations are also relevant for the main
transition from the uniform state into the CDW one at $T = T_c$. We can
think of the primary stripe and bubble phases as examples of a frustrated
phase separation. At $\nu_N = 1 / 2$ the conventional tangent
construction would predict $k_B T_c = E_{\rm ex} / 4$, whereas the
actual (mean-field) transition temperature is lower [Eq.~(\ref{T_c})]
because of the extra energy density associated with the edges of the
stripes.

Returning to the case of superbubbles, we estimate $d^2 f_{s (b)} / d
n^2 \sim f_{s (b)} / n^2 \sim E_{\rm ex} / n^2$ and $\gamma \sim E_{\rm
ex} R_c n_s$ at $T = 0$. The criterion~(\ref{phase_coexistence}) becomes
$(n_s - n_b) / n_s \lesssim R_c \sqrt{n_s} \sim \sqrt{N}$. Thus, we can
be certain that the superstructures {\it do not appear\/} at high LLs.
The cases of small $N$ or high temperatures require further
study~\cite{Fogler_unpub}.

%
%

\subsection{Transitions caused by particle discreteness}

\begin{figure}
\includegraphics[width=2.8in,bb=120 417 447 671]{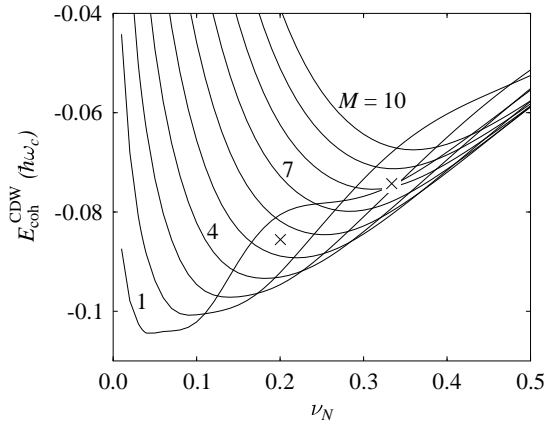}
\caption{
Cohesive energy $E_{\rm coh}^{\rm CDW}$ in a set of bubble phases with
different number of particles per bubble $M$. $E_{\rm coh}^{\rm CDW}$ is
defined as the interaction energy per particle $E_{\rm int}$ relative to
the uncorrelated liquid [where $E_{\rm int} = -(\nu_N / 2) E_{\rm ex}$].
Calculation parameters: $N = 5$, $r_s = \protect\sqrt{2}$, $T = 0$. The
crosses represent the Laughlin liquid energies. (Reproduced with permission
from Fig.~2 of Ref.~\cite{Fogler_97}).
\label{Fig_bubbles}
}
\end{figure}

With the periodicity of the bubble phase set by the preferred wavevector
$q_*$, the area of unit cell of the bubble lattice is equal to $S_0 = 2
\sqrt{3}\,\pi^2 / q_*^2$. The number of particles per bubble is
therefore $M = S_0 \nu_N / 2 \pi l^2$. Using Eq.~(\ref{q_optimal}) we
obtain~\cite{Fogler_96}
\begin{equation}
                   M \approx 3 \nu_N N, \quad N \gg 1.
\label{optimal_bubble}
\end{equation}
It is natural to ask whether this formula should be taken literally,
even when it predicts a nonintegral $M$. Strictly speaking, CDWs with a
fractional number of particles per unit cell are not ruled out. However,
we choose to ignore such an exotic possibility because early
Hartree-Fock studies for the lowest Landau level~\cite{Yoshioka_79}
concluded that only the phases with integer-valued $M$ are stable. In
our own numerical studies only integral $M$ were examined. The results
are reproduced in Fig.~\ref{Fig_bubbles} (see Ref.~\cite{Fogler_97} for
details). We found that the ground state value of $M$ never deviates
from the prediction of Eq.~(\ref{optimal_bubble}) by more than unity at
$N \leq 10$ and all $\nu_N$. This confirms the robustness of the optimal
period $q_*$. It also suggests a practical rule for calculating the
optimal $M$: one should evaluate the right-hand side of
Eq.~(\ref{optimal_bubble}) and round it to the nearest integer.

%
%
Under the assumptions we made, $M$ exhibits a step-like behavior with
unit jumps at a finite set of filling fractions. They correspond to the
first-order transitions between distinct bubble phases. Everything said
earlier about the possible phase co-existence near the first-order
transitions applies here as well. For example, we do not expect any
``bubbles of bubbles'' at large $N$ although at moderate $N$ there is
such a possibility. As $\nu_N$ decreases, the lattice constant of the
bubble phase changes smoothly in between- and discontinuously at the
transitions, but always remains close to $4 \pi / \sqrt{3}\,q_* \approx
3.0 R_c$. Only in the Wigner crystal state ($M = 1$) the lattice
constant is no longer tied to the cyclotron radius but is equal to $(4
\pi / \sqrt{3}\, \nu_N)^{1/2} l$, which is much larger than $R_c$ at
$\nu_N \ll 1 / N$.

One warning is in order here. While very useful, the Hartree-Fock
approximation and the Landau theory of the phase transitions do not
properly account for thermal and quantum fluctuations in 2D. A
preliminary attempt to include these effects reveals additional phases
and phase transitions of different order. The revised phase structure will
be discussed in Sec.~\ref{Liquid_crystals} below.
 
\section{Validity of the Hartree-Fock theory}
\label{HF_validity}

It has been mentioned above that the Hartree-Fock ground state of a
partially filled LL is always a CDW. In particular, for $N = 0$ it is a
Wigner crystal~\cite{Yoshioka_79}. It is well established by now that
the latter prediction is in error at most filling fractions $\nu_N$.
Instead, Laughlin liquids and other fractional quantum Hall states
appear~\cite{FQHE}. The Wigner crystal is realized only when $\nu_N$ is
very small (dilute particles) or very close to $1$ (dilute holes). Below
we present heuristic arguments and numerical calculations, which
together make a convincing case that the situation at large $N$ is {\it
different\/}, so that the ground state has a CDW order at all $\nu_N$
and not just in the dilute limit.

\subsection{Quantum Lindenmann criterion}

A heuristic criterion of the stability of periodic lattices is the
smallness of the fluctuations about the lattice sites compared to the
lattice constant. Unless $r_s$ is extremely small, the stripes and
bubbles are {\it contiguous\/} completely filled (or completely
depleted) regions of the topmost LL. In this situation the fluctuating
objects are the edges of the stripes and bubbles. Using the suitably
modified theory of the quantum Hall edge states (see
Sec.~\ref{Edge_models}), one can estimate their fluctuations $\delta r$
to be of the order of the magnetic length, $\delta r \sim l$. Since the
period $\Lambda$ of the stripe and bubble lattices is at least a few
$R_c = \sqrt{2 N + 1}\,l$, the Lindenmann criterion $\delta r \ll
\Lambda$ is well satisfied, and so the CDW states should be stable at
any $\nu_N$ and sufficiently large $N$. Basically, when the amplitude of
the local filling factor modulation is appreciable and the stripes
(bubbles) are so wide, they are very ``heavy,'' quasiclassical objects
and quantum fluctuations are unable to induce a quantum melting of their
long-range crystalline order. This is in contrast to the small $N$ case
where the quantum fluctuations in the real space are of the order of the
lattice constant except in the dilute limit.

\subsection{Diagrammatic arguments}

In a work~\cite{Moessner_96} published soon after our original
papers~\cite{Fogler_96}, Moessner and Chalker systematically
analyzed the perturbation theory series for the partially filled high
LL problem. They were able to achieve definitive results under two
simplifying assumptions: (a) there is no translational symmetry breaking
and (b) the range $R$ of the quasiparticle interaction is smaller than
$l$. [This interaction is given by the inverse Fourier transform of
$\tilde{v}_0(q) / \epsilon(q)$]. Under such conditions the Hartree-Fock
diagrams were shown to dominate in the $N \gg 1$ limit. This is a very
important result. Even though it enables us to make controlled
statements only about the high-temperature uniform state, it certainly
enhances the credibility of the Hartree-Fock results at all $T$. Due to
the screening by the lower LLs embodied in the dielectric function
$\epsilon(q)$ [Eq.~(\ref{v_q})], the effective range $R$ of interaction
turns out to be of the order of~\cite{Aleiner_95} $a_B = \hbar^2 \kappa
/ m e^2 = l \sqrt{2 / \nu r_s^2}$; thus, condition (b) is satisfied at
$N \gg r_s^{-2}$, which is not too restrictive in practice.

There is still an interesting albeit academic question of what happens
at truly small $r_s$ and intermediate filling factors where $1 \ll N \ll
r_s^{-2}$, so that $R \gg l$. It will be discussed shortly
below.

\subsection{CDW vs Laughlin liquids}

Obvious competitors of the CDW states at simple odd-denominator
fractions $\nu_N = 1 / (2 k + 1)$ are the Laughlin liquids. The
interaction energy per particle in a Laughlin liquid can be found
summing a rapidly converging series~\cite{Girvin_86}
\begin{equation}
 E_{\rm LL} = -\frac{\nu_N}{2} E_{\rm ex}
 + \frac{\nu_N}{\pi} \sum_{K = 1}^\infty c_K V_K,
\label{E_LL}
\end{equation}
where $V_K$ are Haldane's pseudopotentials
\begin{equation}
  V_K = \frac{1}{2\pi} \int\! d^2 q\, \tilde{u}_{\rm H}(q)\,
        F_K(\sqrt{2}\,q),
\end{equation}
$E_{\rm ex}$ (briefly introduced in Sec.~\ref{Quantization} as a
characteristic energy scale) is defined by
\begin{equation}
         E_{\rm ex} = \tilde{u}_{\rm ex}(0) / (2 \pi l^2),
\label{E_ex}
\end{equation}
and $c_K$ are coefficients calculable by the Monte-Carlo
method~\cite{Girvin_86,Fogler_97}. Numerically, about a dozen terms in
the series~(\ref{E_LL}) are needed to get an accurate value of $E_{\rm
LL}$ at $\nu_N = \frac13$ and $\nu_N = \frac15$. In the large-$N$ limit
we can also derive an analytical estimate, guided by the asymptotic
relation
\begin{equation}
  V_K \simeq \tilde{u}_{\rm ex}(2 \sqrt{K} / l) / l^2, \quad K \gg 1.
\label{V_K}
\end{equation}
It indicates that $E_{\rm LL}$ is determined by the behavior of
$\tilde{u}_{\rm ex}(q)$ at $q \sim l^{-1}$. Since the exchange and the
direct interaction potentials are linked by the Fourier transform
[cf.~Eq.~(\ref{u_HF})], $\tilde{u}_{\rm ex}$ has the effective range of
$R / l^2$ in the $q$-space~\cite{Fogler_96}. Thus, two cases have to be
distinguished.

1. $R \ll l$.--- Physically, this corresponds to $N \gg \max\{1,
r_s^{-2}\}$ (recall that the interaction range $R$ is of the order of
$a_B$). Provided $R \ll l$, only first few terms in the
series~(\ref{E_LL}) are important, leading to the estimate $E_{\rm LL}
\sim \tilde{u}_{\rm ex}(1 / l) / l^2$. On the other hand, the
interaction energy per particle in the CDW ground state, $E_{\rm CDW}
\sim -\tilde{u}_{\rm ex}(q_0) / l^2 \sim -E_{\rm ex}$, is significantly
lower, i.e., the CDW wins. In the practical case of $r_s \sim 1$, the
CDW should be lower in energy than the Laughlin liquids at $N \ge N_c$
where $N_c$ is a small number. The numerical calculations reviewed below
indicate that this ``critical'' number is $N_c = 2$.

2. $R \gg l$.--- This regime appears in the parameter window $1 \ll N
\ll r_s^{-2}$. It is mostly of {\it academic\/} interest because it can
be realized only in very high density 2D systems where $r_s \ll 1$. Such
systems are unavailable at present.

The theoretical analysis proceeds as follows. It turns out that
Eq.~(\ref{V_K}) is correct with a relative accuracy $1 / \ln N$ even for
moderate $K$. With the same accuracy we can replace all $V_K$'s in
Eq.~(\ref{E_LL}) by $2 \pi E_{\rm ex}$. Using the sum rule $\sum_K c_K =
(1 - \nu_N^{-1}) / 4$~\cite{Girvin_86}, we arrive at the asymptotic
formula
\begin{equation}
  E_{\rm LL} \simeq -E_{\rm ex} / 2,
\label{E_LL_II}
\end{equation}
which is of the same order as $E_{\rm CDW}$. To compare the energies of
the two states, we have to exercise some care.
Skipping the derivation, which relies heavily on
another sum rule~\cite{Yoshioka_79}, for
the Hartree-Fock states
\begin{equation}
  \textstyle \sum |\langle \tilde{\Delta}({\bf q}) \rangle|^2 = \nu_N,
\label{Sum rule}
\end{equation}
we quote only the final result,
\begin{equation}
                   E_{\rm CDW} \simeq -E_{\rm ex} / 2.
\label{E_CDW}
\end{equation}
It signifies that the Laughlin liquid and the CDW have the same energy
with a relative accuracy of $1 / \ln N$. We may understand this
surprising near equality as follows. Consider a system of particles
interacting via a long-range two-body potential $v(r)$, which is nearly
constant up to a distance $R$ and then gradually decays to zero at
larger distances. The particles are presumed to be spread over a uniform
substrate ``of opposite charge'' with which they interact via a
potential $-n_0 v(r)$. This fixes the average particle density to be
$n_0$. It is easy to see that for any configuration of particles, which
is uniform on the lenghscale of $R$, two potentials nearly cancel each
other: (i) the total potential created at the location of a given
particle by all the other particles of the system and (ii) the potential
due to the substrate. The net potential is equal to $\Sigma = -v(0)$
because the particle does not interacts with itself. Due to the pairwise
nature of the interaction, the interaction energy per particle
(including the interaction with the substrate) is one half of $\Sigma$,
i.e., $-v(0) / 2$. Now we just need to recall from
Sec.~\ref{Quantization} that the energy of the self-interaction in our
system is $E_{\rm ex}$ to recognize Eqs.~(\ref{E_LL_II}) and~(\ref{E_CDW})
as particular cases of this general relation.

The above discussion has several implications. First, it clarifies the
physics behind the arguments of Moessner and Chalker~\cite{Moessner_96}
that the CDW states are likely to face a strong competition
from certain uniform states when the interaction is sufficiently
long-range, $R \gg l$. Second, it leaves the nature of the ground state
at $1 \ll N \ll r_s^{-2}$ an open question at the moment. The CDW states
seem to be favored in numerical calculations done for $2 \leq N \leq 10$
and both $r_s \sim 1$ and $r_s \ll 1$ (see below). Truly high $N$ have
not been investigated yet.

One intriguing possibility is the emergence of novel phases where the
CDW ordering is not static but dynamic. One particular example is a
quantum nematic phase, which can be visualized as a ``soup'' of
fluctuating stripes. Such phases are actively discussed both in the
context of the quantum Hall
effect~\cite{Musaelian_96,Balents_96,Fradkin_99,Fogler_01,Wexler_02,%
Radzihovsky_02} and the high-temperature
superconductivity~\cite{High_Tc_stripes,Fradkin_98,Zaanen_00}. We would
like to reiterate that an experimental search for these exotic phases
would require very special samples, e.g., with $r_s$ much lower than
presently available.

\subsection{Numerical results}

{\it Trial wavefunctions\/}.--- The analytical estimates for $E_{\rm
LL}$ and $E_{\rm CDW}$ derived above become accurate only at very large
$N$. At moderate $N$ the comparison of trial states has to be done
numerically. The procedure of calculating the energies of Laughlin
liquids has been outlined above. It relies on the mapping of the
$N$th LL problem onto a problem at the lowest LL with the modified
interaction, Eq.~(\ref{v_q}). To compute the energy of a Hartree-Fock
CDW state we use the same trick: the trial state is chosen from the
Hilbert space of the lowest LL, but the interaction potential is
appropriately modified.

The first step is to define the wavefunction of a single bubble:
\begin{equation}
\displaystyle \Psi_0\{ {\bf r}_k\}
 = \prod\limits_{1 \leq i < j \leq M} (z_i - z_j) \times
 \exp \left( -\sum\limits_{i = 1}^M \frac{|z_i|^2}{4 l^2}\right),
\label{Bubble_0}
\end{equation}
where $z_j = x_j + i y_j$ are complex coordinates of $M$ quasiparticles
that compose this bubble. A well-known property of the Vandermonde
determinant indicates that $\Psi_0$ is in fact, a Hartree-Fock state. It
is easy to see also that $\Psi_0$ defines the most compact arrangement
of $M$ quasiparticles allowed at the lowest LL, i.e., a circular droplet
of a completely filled LL centered at the point $x = y = 0$.

The trial state we studied is composed of bubbles arranged in triangular
lattice. It can be obtained by replicating the bubble~(\ref{Bubble_0})
and translating its multiple copies to the appropriate lattice sites.
This has to be followed by the antisymmetrization with respect to
particle exchanges among different bubbles. The resulting wavefunction
does not have a simple explicit form. Fortunately, to calculate the
Hartree-Fock energy we do not need the wavefunction but only the
particle density, see Eq.~(\ref{F_HF}) and Ref.~\cite{Yoshioka_79}. An
excellent approximation for the latter is simply the sum of the
densities of the individual bubbles. Strictly speaking, it is not a
fully self-consistent solution of the nonlinear Hartree-Fock equations
because of a small nonorthogonality among the wavefunctions of different
bubbles. However, even the nearest bubbles are effectively so far away
from each other that the deviations from the self-consistency are
extremely small. If desired, the degree of self-consistency can be
further improved using an iterative procedure of the relaxation type
with the described density distribution as the initial guess. We have
done this kind of calculations~\cite{Fogler_97} and found that the
iterations lower the energy of the state by less than one part in
$10^6$, which does not affect the comparison with the Laughlin liquid
energy (known much less accurately).

On the basis of such calculations, we concluded that the CDW becomes the
ground state at $\nu_N = \frac13$ for $N \geq 2$ and at $\nu_N =
\frac15$ for $N \geq 3$. For the latter fraction and $N = 2$ the
energies of the two trial states are so close that no definite
conclusion could be made. Nevertheless, in practice the samples always
contain some amount of disorder, which would favor the CDW state over
the liquid state. Therefore, we established $N_c = 2$ as the
``critical'' LL index where the transition from the Laughlin liquids to
CDW phases occurs. $N_c$ turned out to be the same both for $r_s \sim 1$
and $r_s \ll 1$, and whether or not we included the effect of the
finite-thickness of the 2D layer~\cite{Fogler_97}.

\begin{figure}
\includegraphics[width=3.8in,bb=46 118 544 646]{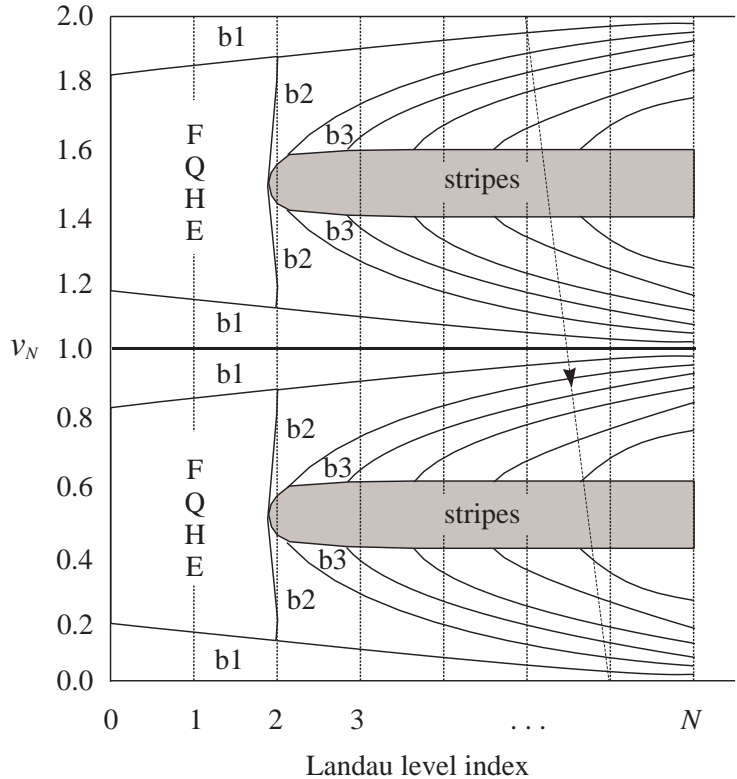}
\caption{
The global phase diagram of a 2D electron system in the axes $(N,
\nu_N)$ for the practical case $r_s \lesssim 1$ (schematically). The
labels ``b1'', ``b2'', {\it etc.\/} denote the bubble phases with 1, 2,
{\it etc.\/} particles per unit cell. As the magnetic field decreases,
the system traces a zigzag path through this diagram, which starts at
the lower left corner and proceeds along continuous segments $(N, 0)
\rightarrow (N, 2)$ connected by discontinuous jumps $(N, 2) \rightarrow
(N + 1, 0)$. One such jump is shown by the dashed line with the arrow.
In principle, points away from the zigzag path can also be sampled if
the functional form of the bare interaction can be modified sufficiently
strongly compared to the Coulomb law.
\label{Fig_global_phase_diagram}
}
\end{figure}

Since the fractional quantum Hall effect (FQHE) is traditionally
associated with the Laughlin states while the CDW does not exhibit the
FQHE, our results imply that the FQHE is restricted to the lowest and
the first excited LLs, $N = 0$ and $N = 1$. {\it It is in principle
impossible to observe the FQHE at\/} $N \geq 2$, and to date no
one has. We are therefore led to propose the global phase diagram of
the 2D electron systems shown in Fig.~\ref{Fig_global_phase_diagram}.

%
%

{\it Exact diagonalization of small systems\/}.--- Strong evidence in
favor of the CDW order in a partially filled $N \geq 2$ LL has been
given by Rezayi, Haldane, and Yang~\cite{Rezayi_99}, who studied systems
up to 14 electrons by means of the direct numerical diagonalization of the
Hamiltonian. Crucial for their success was employing periodical boundary
conditions along the $\hat{\bf x}$ and $\hat{\bf y}$-directions (torus
geometry). This setup avoids imposing the defects into the CDW lattice,
aligns the CDW in a specific direction (which facilitates its
detection), and enables to deduce the number of particles per unit cell
simply from the multiplicity of the ground state manifold.
Rezayi, Haldane, and Yang found no evidence of incompressible FQHE
states at $2 \leq N \leq 6$ and $\nu_N = \frac14, \frac13, \frac25$,
{\it etc\/}. Instead, they reported the ground state degeneracies and
quasi-Bragg peaks in the structure factor fully consistent with the
formation of the stripe phase near the half-filling and a bubble phase
at $\frac13$ and $\frac14$. The periodicity deduced from the quasi-Bragg
peak positions agreed with the Hartree-Fock prediction~(\ref{q_optimal})
within a few percent.

{\it Density matrix renormalization group\/}.--- Shibata
and Yoshioka~\cite{Shibata_01} studied $N = 2$ case using another
powerful numerical technique, the density matrix renormalization group.
Although not exact, it is presumed to be highly accurate both for the
ground state energy and the ground state wavefunction. They were able to
study larger systems, up to 18 electrons. Shibata and Yoshioka
presented pair correlation functions unambiguously showing the stripe
and bubble phases and pinpointed the transition point between them to be
at $\nu_N \approx 0.38$.

\section{Experimental evidence for stripes and bubbles}
\label{Experimental_evidence}

\subsection{Resistance anisotropy}

The existence of the stripe phase as a
physical reality was evidenced by a conspicuous magnetoresistance
anisotropy observed near half-integral fractions of high
LLs~\cite{Lilly_99,Du_99}. This anisotropy develops at low temperatures,
$T \lesssim 0.1\,{\rm K}$, and only in very clean samples. The
anisotropy is the largest at $\nu = 9/2$ ($N = 2$, $\nu_N = 1 / 2$) and
decreases with increasing LL index. At $T = 25\, {\rm mK}$ it remains
discernible up to $\nu \sim 11 \frac12$ whereupon it is washed out,
presumably, due to residual disorder and/or finite temperature. The main
anisotropy axes seem to be always oriented along the crystallographic
axes of the GaAs crystal: $[1 \bar{1} 0]$ (the
high-resistance direction) and $[1 1 0]$ (the
low-resistance one). Especially striking are the data obtained using the
square samples in the van-der-Pauw measurement geometry: at $\nu = 9/2$
the resistances along the two principal directions differ by three
orders of magnitude. However, one has to keep in mind that the
van-der-Pauw measurements exaggerate the bare anisotropy of the
resistivity tensor~\cite{Simon_99}, due to current
channeling along the easy (low-resistance) direction. Indeed, Hall-bar
measurements, which provide a faithful representation of
the resistivity tensor, show much smaller but still a very significant
anisotropy up to 7:1. Once the current distribution effects are taken
into account, the following picture emerges. At $T \gtrsim 0.1\,{\rm
K}$, the transport is isotropic. As $T$ decreases, the resistivity along
the $[1 1 0]$ direction ($\rho_{y y}$) decreases but only
slightly. In contrast, the resistivity in the $[1 \bar{1} 0]$
direction ($\rho_{x x}$) rapidly increases, growing by almost
an order of magnitude when $T$ drops down to $25\,{\rm mK}$. The anisotropy
is the largest at $\nu_N = 1 / 2$ but persists in a sizable interval
$0.4 < \nu_N < 0.6$ of filling factors. Thus, as a function of $\nu_N$,
$\rho_{x x}$ exhibits a peak resembling the quantum Hall
transition peaks in dirtier samples at higher temperatures.

In contrast, the magnetotransport measurements near half-integral
fillings of $N = 0$ and $N = 1$ LLs reveal no significant anisotropies
and no peaks in the longitudinal resistance. Instead, the
resistance exhibits a {\it minimum\/} as a function of
$\nu_N$, which deepens at $T$ decreases~\cite{Pan_Pfaffian_99}. Such
unambiguous distinctions indicate that the electron ground states at
high LLs $N \geq 2$ are qualitatively different from those in
lower LLs. The $\nu = 9 / 2$ is the fraction that demarcates the
transition to the realm of novel high LL physics.

The emergence of the anisotropy is very natural once we assume that the
stripe phase forms. It is based on two concepts: pinning of stripes by
disorder and the edge-state transport. Each of these topics deserves a
separate discussion, which will be given (in a brief form) in
Secs.~\ref{Pinning} and \ref{Edge_models}, respectively. Here we only
sketch the basic ideas. The pinning serves the purpose of preventing the
global sliding of the stripes. This singles out the edge-transport as
the only viable mechanism of current propagation. The edges of the
stripes can be visualized as metallic rivers, along which the transport
is ``easy.'' The charge transfer among different edges, i.e., across the
stripes, requires quantum tunneling and is ``hard'' because the stripes
are effectively far away. Thus, if the stripes are preferentially
oriented along the $[1 1 0]$ direction, the sample would
exhibit the anisotropy of the kind observed in the experiment.

The physical mechanism responsible for the alignment of the stripes
along the definite crystallographic direction is debated at present (see
Sec.~\ref{Other}). In the absence of external aligning fields, the
stripe orientation would be chosen spontaneously. On the other hand, due
to enormous collective response of the stripe-ordered phase, the stripes
can be easily oriented by a tiny bare anisotropy of the medium or the
substrate.

\subsection{New insulating states}

The existence of the {\it bubble phases\/} at high LLs is supported by
another striking experimental discovery: reentrant integral quantum Hall
effect (IQHE) at $\nu \approx 4.25$ and $\nu \approx 4.75$. The Hall
resistance at such filling factors is quantized at the value of the
nearest IQHE plateau, while $\rho_{x x}$ and $\rho_{y y}$ show a deep
minimum with an activated temperature dependence. The transport is
isotropic in these novel insulating states, $\rho_{x x} \approx \rho_{y
y}$. The current-voltage ($I$-$V$) characteristics exhibit pronounced
nonlinearity, switching, and hysteresis. Such phenomena are hallmarks of
the glassy behavior common for pinned crystalline lattices and
conventional CDWs. Hence, these observations are fully consistent with
the theoretical picture of a bubble lattice pinned by disorder. At $N =
2$ we expect only two bubble phases: with two ($M = 2$) and with one ($M
= 1$) particle per bubble. Both phases are subject to pinning and should
be insulating at $T = 0$. To understand the reentrancy phenomenon we
have to take into account the finite-temperature effects. The $M = 2$
bubble phase is more rigid than the $M = 1$ (Wigner crystal) phase, and
remains stable at temperatures where the Wigner crystal is already
melted. As $\nu_N$ is varied towards the nearest integer, the insulating
$M = 2$ bubbles are replaced by a conducting plasma formed in place of
the Wigner crystal. Close enough to the integer $\nu_N$, the plasma is
so dilute that weakly interacting quasiparticles become localized by
disorder and the conventional IQHE results.

Very recently, several more reentrant insulating states has been
discovered also in the $N = 1$ LL~\cite{Eisenstein_xxx}. Their nature
remains to be determined but it is tantalizing to suggest that these are
also the bubbles phases.

\subsection{Other experimental findings}

It was shown in a set of remarkable experiments that the anisotropy near
half-integral fillings can be strongly affected by an in-plane magnetic
field $B_\parallel$. When applied along the easy resistance direction,
the hard and the easy anisotropy axes interchange at $B_\parallel
\gtrsim 0.5\,{\rm T}$~\cite{Pan_99,Lilly_99b}. When applied along the
hard direction, the influence of $B_\parallel$ is much less pronounced
and is to somewhat suppress the anisotropy. This intriguing behavior is
thought to originate from the orbital effects of $B_\parallel$ in a
finite-thickness 2D layer~\cite{Stanescu_00,Jungwirth_99}. They can be
crudely described as squeezing of the cyclotron orbits in the direction
perpendicular to $B_\parallel$. For such distorted orbits the stripe
phase energy depends on the orientation of the stripes with respect to
the in-plane magnetic field. Calculations based on a suitably
generalized Hartree-Fock theory of the previous sections show that the
preferred orientation of the stripes can be both parallel and
perpendicular to the in-plane magnetic field, depending on microscopic
details of the real systems~\cite{Stanescu_00,Jungwirth_99}. For the
specific parameters believed to accurately describe the samples examined
in Refs.~\cite{Pan_99,Lilly_99b}, the perpendicular orientation is
preferred (in agreement with the experiment). However, further
theoretical and experimental work is needed to fully understand these
issues.

The magnetotransport anisotropy at high LLs was also observed for
$p$-type GaAs samples~\cite{Shayegan_00}.

The higher current transport regime near $\nu = 9 / 2$ was investigated.
Gradual increase in the differential resistance along the hard direction
was reported. Compared to the strong nonlinearities at the reentrant
IQHE states, it is a relatively weak effect.

Finally, the degree of anisotropy and the effect of the in-plane fields
were found to be more pronounced in the lower spin subtend of the
same Landau level. A possible explanation within the Hartree-Fock theory
was recently suggested by Wexler and Dorsey~\cite{Wexler_01}.

\section{Many faces of the stripe phase}
\label{Liquid_crystals}

\begin{figure}
\includegraphics[width=3.3in,bb=98 280 534 624]{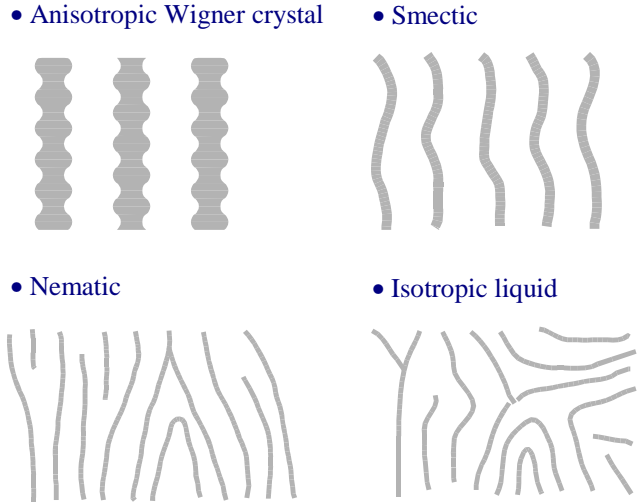}
\caption{
Sketches of possible stripe phases.
\label{Fig_four_phases}
}
\end{figure}

In the wake of the experiments, a considerable amount of work has been
devoted to the stripe phase in recent
years~\cite{Fradkin_99,MacDonald_00,Fertig_99,Fogler_00,Cote_00,Yi_00,%
Barci_01,Lopatnikova_01}. It led to the understanding that the
``stripes'' may appear in several distinct forms: an anisotropic
crystal, a smectic, a nematic, and an isotropic liquid
(Fig.~\ref{Fig_four_phases}). These phases succeed each other in the
order listed as the magnitude of either quantum or thermal fluctuations
increases. Thus, at small $N$ ($N = 2, 3$) or close to $T_c$ the phase
diagrams of Fig.~\ref{Fig_stripes_and_bubbles}a and
Fig.~\ref{Fig_global_phase_diagram} need modifications to incorporate
some (if not all) of those phases. The general structure of the revised
phase diagram for the quantum ($T = 0$) case was discussed in the
important paper of Fradkin and Kivelson~\cite{Fradkin_99}. Pinpointing
the new phase boundaries in terms of the conventional parameters $r_s$
and $\nu$ will require further analytical and numerical work. Once
again, we wish to emphasize that at large $N$ these additional phases
have very narrow regions of existence, if any. Let us now give the
definitions of these intriguing phases and discuss their basic
properties. 

%
%

{\it Stripe crystal\/}.--- This state may in principle be understood at
the Hartree-Fock level. It was shown~\cite{Fertig_99} that the initially
proposed Hartree-Fock solution with smooth edges and a strictly 1D
periodicity~\cite{Fogler_96} is not the global energy minimum. A further
gain in energy is attained once the stripes acquire a periodic
modulation in the longitudinal direction, in antiphase on each pair
of neighboring stripes. The resultant phase breaks the translational
symmetry in {\it both\/} spatial directions and thus is equivalent to a
2D crystal. The unit cell of such a crystal is very anisotropic, with
the aspect ratio of the order of $N : 1$. However, there is only a
single particle per unit cell; thus, this state is an anisotropic Wigner
crystal. It is presumably the true ground state of the system at
sufficiently large $N$ where the Hartree-Fock is deemed to be exact.

{\it Smectic state\/}.--- The usual definition of the smectic is a
``liquid with the 1D periodicity.'' A smectic is less ordered than a
crystal because the translational symmetry is broken only in one spatial
direction. The rotational symmetry is of course broken as well. The
smectic stripe phase can be thought of as a descendant of the stripe
crystal where the longitudinal modulations on neighboring stripes
persist locally, but have no long-range antiphase order because of
dynamic phase slips. In other words, the neighboring stripes are {\it
unlocked\/}~\cite{Fertig_99}. In the thermodynamic limit this kind of
state is equivalent to a stripe phase with no modulations (smooth
edges), akin to the original Hartree-Fock solution~\cite{Fogler_96}. The
necessary condition for the smectic order is the continuity of the
stripes. If the stripes are allowed to rupture, the dislocations are
created. They destroy the 1D positional order and convert the smectic
into the nematic.

{\it Nematic state\/}.--- By definition, the nematic is an anisotropic
liquid. There is no long-range positional order. As for the
orientational order, it is long-range at $T = 0$ and quasi-long-range
(power-law correlations) at finite $T$. The nematic is riddled with
dynamic dislocations. Other types of topological defects, the
disclinations, may also be present but remain bound in pairs, much
like vortices in the 2D $X$-$Y$ model.

{\it Isotropic liquid\/}.--- Once the disclinations in the nematic
unbind, all the spatial symmetries are restored. The resultant state is
an isotropic liquid with short-range stripe correlations. As the
fluctuations due to temperature or quantum mechanics increase further,
it gradually crosses over to the ``uncorrelated liquid'' where
even the local stripe order is obliterated.

\section{Effective theories of the stripe phase}
\label{Effective_theory}

As often the case, the low-frequency long-wavelength physics of the
system is governed by an effective theory involving a relatively small
number of dynamical variables. The basic form of the effective theory is
essentially fixed by the symmetry considerations. Let us outline how
such theories are constructed for the various stripe states introduced
in the previous section.

\subsection{Stripe crystal}

The low-energy dynamical variables of this state are the elastic
deformation ${\bf u} = \{u_x({\bf r}, t), u_y({\bf r}, t)\}$ of the
crystalline lattice and the effective description is basically the
elasticity theory. In addition, we have to account for the long-range
Coulomb interaction between the density fluctuations $n({\bf r}, t) =
-n_0 \nabla {\bf u}$, where $n_0 = \nu_N / (2 \pi l^2)$ is the average
particle density at the $N$th LL. The symmetry arguments identify four
nonvanishing elastic moduli: $c_{11}$, $c_{12}$, $c_{22}$, and $c_{44}$,
so that the effective Hamiltonian takes the form
\begin{eqnarray}
H &=& \frac{c_{11}}{2} (\partial_x u_x)^2
  + \frac{c_{22}}{2} (\partial_y u_y)^2
  + c_{12} \partial_x u_y  \partial_y u_x
  + \frac{c_{44}}{8} (\partial_x u_y + \partial_y u_x)^2
\nonumber\\
  &+& \frac12 n_0^2 (\nabla {\bf u}) u_{\rm H} (\nabla {\bf u}),
\label{H_crystal}
\end{eqnarray}
where $u_{\rm H}$ should be understood as the
integral operator. The dynamics of the system is governed by the
Lorentz force and can be studied with the help of the
effective Lagrangean
\begin{equation}
{\cal L} =  m n_0 \omega_c u_y \partial_t u_x - H.
\label{L_crystal}
\end{equation}
Solving the corresponding equations of motion, we find the following
excitation spectrum of lattice vibrations (magnetophonons):
\begin{equation}
\omega({\bf q}) =  \frac{\omega_p(q)}{\omega_c} q
\left[\frac{c_{44} + (c_{11} + c_{22} - 2 c_{12} - c_{44}) \sin^2 2\theta}
           {4 m n_0}\right]^{1/2}.
\label{omega_crystal}
\end{equation}
Here $\omega_p(q) = [n_0 \tilde{u}_{\rm H}(q) q^2 / m]^{1/2}$ is the
plasma frequency and $\theta = \arctan (q_y / q_x)$ is the angle between
the propagation direction and the $\hat{\bf x}$-axis. For Coulomb
interactions $\omega_p(q) \propto \sqrt{q}$ leading to the well-known
dispersion relation~\cite{Bonsall_77} $\omega(q) \propto q^{3 / 2}$. In
a strongly anisotropic stripe crystal, $c_{12}, c_{44} \ll c_{11}$
causing the angular dependence $\omega({\bf q}) \propto \sin 2 \theta$
starting from relatively small $q$. All these results are valid for an
idealized clean system. In reality the low-$q$ magnetophonon modes will
be profoundly affected by disorder, which will be discussed in
Sec.~\ref{Pinning}.

\subsection{Smectic state}

As mentioned above, the smectic state is most closely related to the
original Hartree-Fock solution~\cite{Fogler_96}. It may be visualized as
Hartree-Fock stripes slightly decorrelated by phonon-like thermal
fluctuations.

{\it Harmonic approximation\/}.--- 
In the smectic only $u_x$ retains its direct meaning of the elastic
displacement. On the other hand, the density fluctuations $n$ become an
independent degree of freedom. For example, in the case of
incompressible stripes, $n$ comes from the stripe width fluctuations,
which are separate from the ``shape'' fluctuations described by $u_x$.
The number of dynamical variables in the smectic and in the crystal is
therefore the same. Moreover, the smectic can be thought of as a crystal
that lost its shear rigidity because of phase slips between nearby
crystalline rows. This intuitive picture enables us to deduce the
effective theory for the smectic from Eqs.~(\ref{H_crystal}) and
(\ref{L_crystal}) by a certain reduction. Of course, at the end we
should verify that we did not miss any terms allowed by symmetry. The
first step is to formally reintroduce $u_y$ as a solution of the
equation $\partial_y u_y = -n / n_0 - \partial_x u_x$ and use it to
replace all instances of $\partial_y u_y$ in Eq.~(\ref{H_crystal}). To
ensure that the stripes are free to slide with respect to each other in
the $\hat{\bf y}$-direction, the terms that depend on $\partial_x u_y$
should be dropped: $c_{12} \rightarrow 0$, $c_{44} \rightarrow 0$. Yet
we need to be careful and recall that the complete elasticity theory
always contains higher gradients such as $(\partial_y^2 u_x)^2$
[suppressed in Eq.~(\ref{H_crystal})]. Once the coefficient in front of
the first-order gradient term $(\partial_y u_x)^2$ vanishes, higher
gradients become dominant and must be included. The resultant effective
Hamiltonian and the Lagrangean take the form
\begin{eqnarray}
&& H = \frac{Y}{2} (\partial_x u)^2 + \frac{K}{2} (\partial_y^2 u)^2
+ \frac12 n (u_{\rm H} + \chi^{-1}) n + C n \partial_x u,
\label{H_smectic}\\
&& {\cal L} =  p \partial_t u - H,\quad
  \partial_y p = -m \omega_c (n + n_0 \partial_x u).
\label{L_smectic}
\end{eqnarray}
Here we switched to notations more natural for the smectic: $u_x$ became
$u$, $u_y$ was traded for the canonical momentum $p$, the sum $c_{11} +
c_{22}$ became $Y$, {\it etc\/}. The physical meaning of new
phenomenological coefficients is as follows. $Y$ and $K$ are the
compression and the bending elastic moduli, $\chi$ is the
compressibility, and $C = 2 \pi l^2 Y d \ln \Lambda / d \nu_N$ accounts
for the dependence of the mean interstripe separation $\Lambda$ on the
average filling factor.


Since the number of dynamical variables in the smectic is the same as in
the crystal state, the collective mode count is also unchanged. We will
keep referring to them as magnetophonons. Solving the equations of
motion for $n$ and $u$ we obtain the dispersion relation of such
magnetophonons~\cite{Fogler_00}:
\begin{equation}
\omega({\bf q}) =  \frac{\omega_p(q)}{\omega_c} \frac{q_y}{q}
\left[\frac{Y q_x^2 + K q_y^4}{m n_0}\right]^{1/2}.
\label{omega_smectic}
\end{equation}
Unless propagate nearly parallel to the stripes, $\omega({\bf q})$ is
proportional to $\sin 2 \theta\, q^{3 / 2}$. Unlike in the stripe crystal,
this relation is obeyed even at $q \rightarrow 0$ (again, in the absence
of disorder or orienting fields). One immediate consequence of this
dispersion is that the largest velocity of propagation for the
magnetophonons with a given $q$ is achieved when $\theta = 45^\circ$.

{\it Thermal fluctuations and anharmonisms\/}.--- From
Eq.~(\ref{H_smectic}) we can readily calculate the mean-square
fluctuations of the stripe positions at finite $T$, e.g.,
\begin{equation}
\langle [u(0, 0) - u(0, y)]^2 \rangle = 
2 \int \frac{d^2 q}{(2 \pi)^2} \frac{k_B T}{Y q_x^2 + K q_y^4}
(1 - e^{i k_y y}) = \frac{k_B T}{2 \sqrt{Y K}} |y|.
\label{uu_smectic}
\end{equation}
This formula is valid if $y$ is large so that magnetophonons with
wavevectors $q_y \lesssim 1 / y$ can be treated classically [$\hbar
\omega({\bf q}) \ll k_B T$]. As one can see, at any finite temperature
magnetophonon fluctuations are growing without a bound; hence, the
positional order of a 2D smectic is totally
destroyed~\cite{DeGennes_book} at sufficiently large distances along the
$\hat{\bf y}$-direction, $|y| \gg \Lambda \sqrt{Y K} / k_B T \equiv
\xi_y$. Similarly, along the $\hat{\bf x}$-direction, the positional
order is lost at lengthscales larger than $\xi_x = (Y / K)^{1/2} \xi_y^2$.

\begin{figure}
\includegraphics[width=2.9in,bb=131 367 499 535]{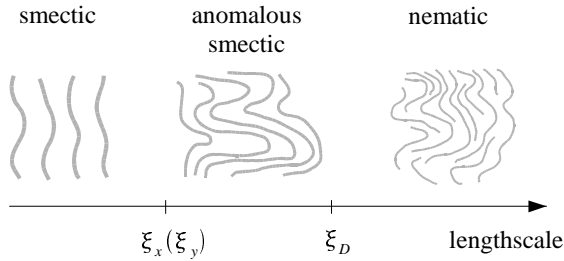}
\caption{
Portraits of the stripe phase on different lengthscales.
\label{Fig_spaghetti}
}
\end{figure}

Another type of excitations, which decorrelate the stripe positions are
the aforementioned dislocations. The dislocations in a 2D smectic have a
finite energy $E_D \sim K$. At $k_B T \ll E_D$ the density of thermally
excited dislocations is of the order of $\exp(-E_D / k_B T)$ and the
average distance between dislocations is $\xi_D \sim \Lambda \exp(2 k_B
T / E_D)$. At low temperatures $\xi_x, \xi_y \ll \xi_D$; therefore, the
following interesting situation emerges (Fig.~\ref{Fig_spaghetti}). On
the lengthscales smaller than $\xi_y$ (or $\xi_x$, whichever
appropriate) the system behaves like a usual smectic where
Eqs.~(\ref{H_smectic}--\ref{omega_smectic}) apply. On the lengthscales
exceeding $\xi_D$ it behaves\footnote{In a more precise
treatment~\cite{Golubovic_92}, the lengthscales $\xi_{D x} \propto
\xi_D^{6/5}$ and $\xi_{D y} \propto \xi_D^{4/5}$ are introduced such
that $\xi_{D x} \xi_{D y} = \xi_D^2$.} like a nematic~\cite{Toner_81}.
In between the system is a smectic but with very unusual properties. It
is topologically ordered (no dislocations) but possesses enormous
fluctuations. In these circumstances the harmonic elastic theory becomes
inadequate and anharmonic terms must be included. The most important
anharmonisms are captured in the following elastic Hamiltonian, which
should be substituted in place of the first two terms in
Eq.~(\ref{H_smectic})~\cite{DeGennes_book}:
\begin{equation}
H_{el} = \frac{Y}{2} \Bigl[\partial_x u - \frac12 (\nabla u)^2\Bigr]^2
       + \frac{K}{2} (\partial_y^2 u)^2.
\label{H_smectic_el}
\end{equation}
It can be easily checked that the expression inside the square brackets,
which is the compressional strain, is invariant under rotations of the
reference frame by an {\it arbitrary\/} angle $\phi$. For example, if in
the initial reference frame $u = 0$, then in the new frame $u(x, y) = (1
- \cos\phi) x + \sin\phi\, y$ so that $\partial_x u - \frac12 (\nabla
u)^2 = 0$.

%
%

What is the role of anharmonisms? As shown by Golubovi\'c and
Wang~\cite{Golubovic_92}, they cause power-law dependence of the
parameters of the effective theory on the wavevector ${\bf q}$:
\begin{equation}
 Y \sim Y_0 (\xi_y q_y)^{1/2},\quad  K \sim K_0 (\xi_y q_y)^{-1/2},
\label{limit_A}
\end{equation}
for $q_x \ll \xi_x^{-1} (q_y \xi_y)^{3/2}$, $q_y \ll \xi_y^{-1}$, and
\begin{equation}
 Y \sim Y_0 (\xi_x q_x)^{1/3},\quad  K \sim K_0 (\xi_x q_x)^{-1/3},
\label{limit_B}
\end{equation}
for $q_x \ll \xi_x^{-1}$ and $q_y \ll \xi_y^{-1} (q_x \xi_x)^{2/3}$. The
lengthscale dependence of the parameters of the effective theory is a
common feature of fluctuation-dominated phenomena. Other famous examples
include the criticality near phase transitions and in systems at their
lower critical dimension, such as the 2D $X$-$Y$ model. It should be
mentioned that the lower critical dimension for the smectic order is $d
= 3$~\cite{DeGennes_book}, so that the 2D smectic is {\it below\/} its
lower critical dimension. This is the reason why the scaling
behavior~(\ref{limit_A}) and (\ref{limit_B}) does not persist
indefinitely but eventually breaks down above the lengthscale $\xi_D$
where the crossover to the thermodynamic limit of the nematic behavior
commences.

Equations~(\ref{limit_A}) and (\ref{limit_B}) indicate that the
compression modulus decreases while the bending modulus increases as the
lengthscale grows. This can be understood from the following qualitative
reasoning. Thermal fluctuations create a lot of wiggles on the stripes.
When an external compressional stress is applied, it can be
relieved not just by compression but by flattening of the stripes. Since
the latter involves unbending of the crumpled stripes and the bending
costs less energy than the compression ($q^4$ instead of $q^2$), the
apparent compressional modulus $Y$ is smaller than its bare value $Y_0$.
Similarly, when one attempts to bend crumpled stripes, some compression
is necessarily involved, and so the bending modulus $K$ appears larger.

The scaling shows up not only in the static properties such as $Y$ and
$K$ but also in the dynamics. The role of anharmonisms in the dynamics
of conventional 3D smectics has been investigated by Mazenko {\it et
al.\/}~\cite{Mazenko_83} and also by Kats and
Lebedev~\cite{Kats_Lebedev}. For the quantum Hall stripes the analysis
had to be done anew because here the dynamics is totally different. It
is dominated by the Lorentz force rather than a viscous relaxation in
the conventional smectics. This task was accomplished in
Ref.~\cite{Fogler_00}. The calculation was based on the
Martin-Siggia-Rose formalism combined with the $\epsilon$-expansion
below $d = 3$ dimensions. One set of results concerns the spectrum of
the magnetophonon modes, which becomes
\begin{equation}
\omega({\bf q}) \sim \sin \theta \cos^{7/6}\theta\,
(\xi_x q)^{5/3} \frac{\omega_p(\xi_x^{-1})}{\omega_c \xi_x}
\sqrt{\frac{Y_0}{m n_0}}.
\label{omega_m_R}
\end{equation}
Compared to the predictions of the harmonic theory,
Eq.~(\ref{omega_smectic}), the $q^{3/2}$-dispersion changes to
$q^{5/3}$. Also, the maximum propagation velocity is achieved for the
angle $\theta \approx 53^\circ$ instead of $\theta = 45^\circ$. These
modifications, which take place at long wavelengths, are mainly due to
the renormalization of $Y$ in the static limit and can be obtained
by combining Eqs.~(\ref{omega_smectic}) and (\ref{limit_B}). Less obvious
dynamical effects peculiar to the quantum Hall smectics include a novel
dynamical scaling of $Y$ and $K$ as a function of frequency and a
specific $q$-dependence of the magnetophonon damping~\cite{Fogler_00}.

The latter issue touches on an important point. Our effective theory
defined by Eqs.~(\ref{H_smectic}) and (\ref{L_smectic}) is based on the
assumption that $u$ and $n$ are the only low-energy degrees of freedom.
It is probably well justified at $T \rightarrow 0$ but becomes incorrect
at higher temperatures. The point of view taken in Ref.~\cite{Fogler_00}
is that in the latter case thermally excited quasiparticles (``normal
fluid'') should appear and that they should bring dissipation into the
dynamics of the magnetophonons.

Another intriguing possibility is for quasiparticles or other additional
low-energy degrees of freedom to exist even at $T = 0$. Such more
complicated smectic states are not ruled out and are interesting
subjects for future study.

\subsection{Nematic state}

\begin{figure}
\includegraphics[width=2.0in,bb=128 526 478 657]{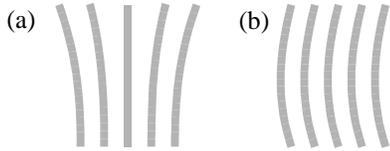}
\caption{
Splay (a) and bend (b) distortions in a nematic.
\label{Fig_splay_bend}
}
\end{figure}

Much like loosing the shear rigidity due to phase slips converts a
crystal to a smectic, loosing the compressional rigidity due to mobile
dislocations can convert a 2D smectic into a nematic. The collective
degree of freedom associated with the nematic ordering is the angle
$\phi({\bf r}, t)$ between the local normal to the stripes ${\bf N}$ and
the $\hat{\bf x}$-axis orientation. Due to inversion symmetry, ${\bf N}$
and $-{\bf N}$, i.e., $\phi$ and $\phi + \pi$ are equivalent, that is
why ${\bf N}$ is often referred to as the {\it director\/} rather than a
vector~\cite{DeGennes_book}. The effective Hamiltonian for ${\bf N}$ is
dictated by symmetry to be
\begin{equation}
H_N = \frac{K_1}{2} (\nabla {\bf N})^2
  + \frac{K_3}{2} |\nabla \times {\bf N}|^2.
\label{H_nematic}
\end{equation}
The coefficients $K_1$ and $K_3$ are termed the splay and the bend Frank
constants~\cite{DeGennes_book}. They control the cost of the two
possible elementary types of director nonuniformity shown in
Fig.~\ref{Fig_splay_bend}. Note that in the smectic phase $\phi =
-\partial_y u$. This entails the relation $K_3 \simeq K$ between the
parameters of the nematic and its parent smectic. On the other
hand, the value of $K_1$ is expected to be determined largely by the
properties of the dislocations~\cite{Toner_81}. The elastic part has a
particularly simple form if $K_1 = K_3$, in which case $H_N = (K_3 / 2)
(\nabla \phi)^2$ just like in the $X$-$Y$ model.

%
%

Another obvious degree of freedom in the nematic are the density
fluctuations $n({\bf r}, t)$. A peculiar fact is that in the static
limit $n$ is totally decoupled from ${\bf N}$, and so it does not enter
Eq.~(\ref{H_nematic}). However, since the nematic is less ordered than
even a smectic, the question about extra low-energy degrees of freedom
or additional quasiparticles become especially relevant. We believe that
different types of quantum Hall nematics are possible in nature. In the
simplest case scenario ${\bf N}$ and $n$ are the only low-energy degrees
of freedom. This type of state has been studied by
Balents~\cite{Balents_96} and recently by the present
author~\cite{Fogler_01}. It was essentially postulated that the
effective Largangean takes the form 
\begin{equation}
{\cal L} = \frac12 \gamma^{-1} (\partial_t {\bf N})^2 - H.
\label{L_nematic}
\end{equation}
(As hinted above, the full expression contains also couplings between
$\partial_t {\bf N}$ and mass currents but they become vanishingly small
in the long-wavelength limit). The collective excitations
are charge-neutral fluctuations of the director. They have a
linear dispersion,
\begin{equation}
\omega({\bf q}) =  q \sqrt{{K_1}{\gamma} \cos^2 \theta +
                           {K_3}{\gamma} \sin^2 \theta},
\label{omega_nematic}
\end{equation}
and resemble spinwaves in the $X$-$Y$ quantum rotor model.

Also discussed in Ref.~\cite{Fogler_01} was a dislocation-mediated
mechanism of the smectic-nematic transition within the framework of
duality transformations developed earlier by Toner and
Nelson~\cite{Toner_81}, Toner~\cite{Toner_82}, and Fisher and
Lee~\cite{Fisher_89}. This theory predicts the existence of a second
excitation branch with a small gap. This gapped mode can be considered a
descendant of the magnetophonon mode of the parent smectic.

Very recently, Radzihovsky and Dorsey~\cite{Radzihovsky_02} formulated a
different theory of the quantum Hall nematics, whose predictions
disagree with our Eqs.~(\ref{L_nematic}) and (\ref{omega_nematic}).
Logically, there are two possibilities. Either, as mentioned above,
there are several distinct kinds of nematic states possible in nature or
some of the theoretical constructions advanced in
Refs.~\cite{Balents_96,Fogler_01,Radzihovsky_02} are incorrect. To
resolve these issues it is imperative to bring the discussion from the
level of effective theory to the level of quantitative calculations. One
promising direction is to investigate concrete trial wavefunctions of
quantum nematics, e.g., the one proposed by Musaelian and
Joynt~\cite{Musaelian_96}:
\begin{equation}
\Psi = \prod\limits_{j < k} (z_{j} - z_{k}) [(z_{j} - z_{k})^2 - a^2]
\times \exp \Bigl(-\sum_j |z_j|^2 / 4 l^2\Bigr).
\label{nematic_trial_wavefunction}
\end{equation}
Here $a$ is a complex parameter that determines the degree of
orientational order and the direction of the stripes. This
particular wavefunction corresponds to $\nu = \frac13$. (As explained
earlier, it can also be used investigate the higher Landau level states
with $\nu_N = \frac13$). Recently, the work in this direction was
continued by Ciftja and Wexler~\cite{Wexler_02}.

Other contributions to the theory of quantum Hall nematics have been
focused on a finite temperature case. They include a work of
Fradkin~{\it et al\/}~\cite{Fradkin_00} who investigated the role of
external anisotropy field in the 2D $X$-$Y$ formulation and also a paper
by Wexler and Dorsey~\cite{Wexler_01}, where a quantitative Hartree-Fock
analysis of the Toner-Nelson disclination unbinding scenario has been
done.

It is worth mentioning that in the quantum case the chain of transitions
crystal $\rightarrow$ smectic $\rightarrow$ nematic $\rightarrow$
isotropic liquid may or may not be realized in full. A finite-size study
by Rezayi {\it et al.\/}~\cite{Rezayi_00} suggests that the transition
from the smectic to an isotropic phase, as the interaction parameters
are varied away from their $N = 2$ Coulomb values, can also occur via a
first-order transition, without the intermediate nematic phase. The
resultant isotropic state is highly correlated and has little in common
with the ``uncorrelated'' Hartree-Fock liquid. The natural candidates
for the isotropic state include a Fermi-liquid-like state of composite
fermions and a Pfaffian (Moore-Read) state~\cite{Rezayi_00}.
Understanding all these competing quantum orders and transitions between
them remains a major intellectual challenge. In contrast, the phase
structure at larger $N$, away from the ``transition point'' $N = 2$, is
much simpler and is adequately captured by the Hartree-Fock theory.

\section{Edge state models}
\label{Edge_models}

Another theoretical approach to the physics of the stripe phases is
based on the edge-state formalism. The advantages of the edge-state
models are two-fold. First, they allow to bridge the gap between the
microscopic theory and the formulations based on the elasticity theory
or hydrodynamics (albeit under some crucial simplifying assumptions).
Second, they enable one to calculate not only collective but also
single-particle properties, such as the tunneling density of states. On
the other hand, being more specialized than the hydrodynamics, the edge
models have a somewhat restricted domain of validity. The models
considered so
far~\cite{Fradkin_99,MacDonald_00,Barci_01,Lopatnikova_01}
apply when the stripes are incompressible. i.e., when they are
contiguous regions of $\nu_N = 1$ separated by {\it sharp\/} boundaries
from the regions with $\nu_N = 0$. In our opinion the edge state models
are better suited for describing the crystal and the smectic states. In
the nematic state stripes exhibit violent quantum (or thermal)
fluctuations and the picture of well defined sharp edges is
questionable.

%
%
\begin{figure}
\includegraphics[width=3.0in,bb=126 403 476 587]{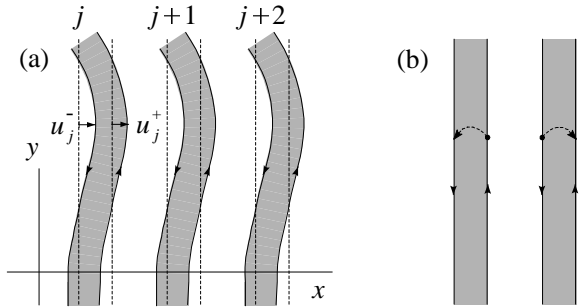}
\caption{
(a) Geometrical meaning of variables in the edge-state effective theory.
(b) Backscattering.
\label{Fig_edges}
}
\end{figure}

The first step in constructing the edge theory is to identify the
low-energy degrees of freedom with the deviations $u^j_\pm$ of the
stripe boundaries from their equilibrium positions $x^j_\pm = (j \pm
\nu_N / 2) \Lambda$, see Fig.~\ref{Fig_edges}a. Such deviations induce
extra 1D guiding center densities along the edges, $\rho^j_\pm = \pm
u^j_\pm / (2 \pi l^2)$. The next step is to find an effective
Hamiltonian $H$ that governs the interaction among $\rho^j_\pm$. To this
end it is convenient to introduce the Fourier transform
\begin{equation}
  \tilde\rho_\sigma({\bf q}) = \Lambda \sum_j \exp (-i q_y x^j_\sigma)
  \int d y e^{-i q_y y} \rho_\sigma^j(y),
  \label{rho_q}
\end{equation}
and define the center-of-mass stripe displacement $u_x$ and the net
local density fluctuation per unit area,
\begin{eqnarray}
&\tilde{u}_x({\bf q}) = 2 \pi l^2 \left[
              \exp(\frac{i}{2} q_x \nu_N \Lambda) \tilde\rho_+({\bf q})
           +  \exp(-\frac{i}{2} q_x \nu_N \Lambda) \tilde\rho_-({\bf q})
\right],&
\label{u}\\
&\tilde{n}({\bf q}) = \Lambda^{-1} \left[
           \exp(\frac{i}{2} q_x \nu_N \Lambda) \tilde\rho_+({\bf q})
        -  \exp(-\frac{i}{2} q_x \nu_N \Lambda) \tilde\rho_-({\bf q})
\right].&
\label{n}
\end{eqnarray}
The general symmetry requirements compel the long-wavelength part $H_0$
of the Hamiltonian to take the form~(\ref{H_smectic}), reproduced here
for convenience:
\begin{equation}
 H_0 = \frac{Y}{2} (\partial_x u_x)^2 + \frac{K}{2} (\partial_y^2 u_x)^2
+ \frac12 n (u_{\rm H} + \chi^{-1}) n + C n \partial_x u_x.
\label{H_smectic_II}
\end{equation}
The edge-state formalism enables one go further and obtain quantitative
estimates for the phenomenological parameters $Y$, $K$, $\chi$, and $C$
from the microscopic Hamiltonian. It should be clarified that the exact
calculation of these parameters remains out of reach for the moment.
Available
derivations~\cite{MacDonald_00,Fertig_99,Yi_00,Wexler_01,Lopatnikova_01}
are certainly not exact. It can be shown~\cite{Fogler_unpub} that all of
them rely on one specific approximation scheme, the time-dependent
Hartree-Fock approximation (TDHFA). In the spirit of the discussion in
the previous sections, we expect the TDHFA to be quantitatively accurate
at $N \gg 1$ but only qualitatively correct in the case of current
experimental interest, $N \sim 2$. Let us therefore proceed
with the exposition of the field-theoretic edge-state formalism.

Besides $H_0$, the full effective Hamiltonian contains another term
$H_{2 q_F}$, whose importance was first emphasized by Fradkin and
Kivelson~\cite{Fradkin_99}. To clarify its origin let us recall that in
the Landau gauge $A = (0, B x, 0)$ each single-particle basis state
$|X\rangle$ has a definite momentum in the $\hat{\bf y}$-direction,
proportional to the mean value $X$ of the $x$-coordinate in such a
state, $k_y = X / l^2$. In the simplest mean-field realization of the
stripes, a given state is filled if $k_y \in (x^j_+ / l^2, x^j_- / l^2)$
and is empty otherwise. Thus, $k_y = x^j_\pm$ play the role of Fermi
points in an equivalent quasi-1D electron system. Note that ``$+$''
(``$-$'') stands for the sign of the Fermi velocity near a given Fermi
points, in the positive (negative) $\hat{\bf y}$-direction. The density
fluctuations $\rho^j_\pm$ propagate along the edges only in the
direction specified by this sign. Such a property is called chirality.
Thus, the stripe phase is characterized by a Manhattan grid of
alternating {\it chiral\/} edge states. From this perspective, $H_0$
describes the particle-particle interactions, which cause only small
changes in $X$ and thus small momentum transfer in the vicinity of the
Fermi points. Another low-energy process is to scatter particles between
different Fermi points, e.g., $x^j_+ / l^2 \rightarrow x^j_- / l^2$ and
$x^{j + 1}_- / l^2 \rightarrow x^{j + 1}_+ / l^2$, see
Fig.~\ref{Fig_edges}b. Such scattering acts must involve a pair of
particles to conserve the total momentum: one particle gains and the
other looses the same amount of momentum, $2 q_F^e \equiv (x^j_+ - x^j_-)
/ l^2 = \nu_N \Lambda / l^2$. $H_{2 q_F}$ accounts precisely for the
processes of this type. Since they cause the reversal of the propagation
direction for the particles involved, it is natural to call it
backscattering. Remarkably, it is possible~\cite{Wen_92} to express
$H_{2 q_F}$ in terms of density fluctuations $\rho^j_\pm$. Without going
into details, we quote the final result~\cite{Fradkin_99,MacDonald_00},
\begin{eqnarray}
H_{2 q_F} &=& -\sum_j \lambda^e \cos[(\phi^j_+ - \phi^j_-)
            + (\phi^{j + 1}_- - \phi^{j + 1}_+)]
\nonumber\\
 &-& \sum_j \lambda^h \cos[(\phi^j_- - \phi^{j - 1}_+)
            + (\phi^j_+ - \phi^{j + 1}_-)]
\label{H_2q_F}
\end{eqnarray}
Here $\phi^j_\pm$ are auxiliary dynamical variables
related to $\rho^j_\pm$ as follows:
\begin{equation}
           \rho^j_\pm = \partial_y \phi^j_\pm / (2 \pi).
\label{phi}
\end{equation}
The phenomenological parameters $\lambda^a$ are proportional to
$\tilde{u}_{\rm H}(2 q_F^a)$ but in general depends on how the theory is
formulated (the ultraviolet cutoff). In principle, interactions can
scatter the particles not only between nearest edges but also
next-nearest ones, {\it etc\/}. The corresponding amplitudes are
proportional to $\tilde{u}_{\rm H}(q)$ where $q$ is the appropriate
momentum transfer. Since $\tilde{u}_{\rm H}$ decreases exponentially at
such large $q$, these other processes are expected to have negligible
effect. Thus, the total Hamiltonian is $H = H_0 + H_{2 q_F}$.
 
The final step in constructing the effective edge theory is determining
the kinetic term. Either from Wen's bosonization theory~\cite{Wen_92} of
by comparing Eqs.~(\ref{L_smectic}) and (\ref{phi}), one can come to the
conclusion that $\phi^j_\pm$ and $\rho^j_\pm$ are canonically conjugate
variables, so that the appropriate Lagrangean is
\begin{equation}
{\cal L} = \frac{\hbar}{4 \pi} \sum_{j, \sigma = \pm}
\sigma \partial_t \phi^j_\sigma \partial_y \phi^j_\sigma - H_0
 - H_{2 q_F}.
\label{L_stripes}
\end{equation}
Let us now explain how this theory can lead either to crystalline or to
smectic behavior. The idea is to treat $H_{2 q_F}$ as a small
perturbation~\cite{Fradkin_99,MacDonald_00,Fertig_99,Yi_00}. If this
perturbation is {\it irrelevant\/}, in the long-wavelength low-frequency
limit $H$ reduces to $H_0$ and ${\cal L}$ to the smectic
form~(\ref{L_smectic}). On the other hand, if $H_{2 q_F}$ is relevant,
then the ``$+$'' and ``$-$'' edges of each stripe become strongly mixed
so that a static $2 q_F$ density modulation in the $\hat{\bf
y}$-direction appears. Its spatial period is $a = 2 \pi / (2 q_F) = 2
\pi l^2 / \nu_N \Lambda$; hence, the number of particles on a given
stripe within one modulation period is $\nu_N \Lambda \times a / (2 \pi
l^2) = 1$. This can be visualized as a 1D chain of particles. The
neighboring chains are locked in antiphase to lower their interaction
energy. This state is an anisotropic Wigner crystal with the aspect
ratio of the unit cell $\Lambda : a = (\nu_N / 2 \pi) (\Lambda / l)^2
\sim N : 1$. Two types of stripe crystals are possible. If the first term
in $H_{2 q_F}$ is relevant, it is an ``electron'' crystal, if the
second term is relevant, it is a ``hole'' crystal.

The elasticity theory~(\ref{H_crystal}) of, e.g., the electron crystal
can be recovered as follows. Since the first sum in Eq.~(\ref{H_2q_F})
is relevant, the large fluctuations of the argument of the corresponding
cosines are inhibited. It is legitimate to expand these cosines to
arrive at
\begin{eqnarray}
  && H_{2q_F} \rightarrow -\sum_j \frac{\lambda_*}{2}
                       (u^j_y - u^{j + 1}_y - \Lambda \partial_y u_x)^2,
\label{H_2q_F_expanded}\\
  && u_y({\bf q}) \equiv \frac{l^2}{\nu_N \Lambda}
  [\phi_+({\bf q}) e^{i \Lambda q_x / 2}
 - \phi_-({\bf q}) e^{-i \Lambda q_x / 2}].
\label{u_y}
\end{eqnarray}
In the long-wavelength limit it gives a term proportional to
$(\partial_x u_y + \partial_y u_x)^2$, which makes the shear modulus
$c_{44}$ finite [cf.~Eq.~(\ref{H_crystal})]. A more careful treatment
presumably recovers the remaining elastic constant $c_{12}$. Our choice
of $u_y$ in Eq.~(\ref{u_y}) is not arbitrary because we want to preserve
the physical interpretation of $u_y$ as the elastic displacement. This
requires $H_{2 q_F}$ to be invariant under the shift by a lattice
constant, $u_y \rightarrow u_y + a$. Since $H_{2 q_F}$ is invariant
under the shift $\phi^j_\pm \rightarrow \phi^j_\pm + 2\pi$, it fixes the
coefficient of proportionality in Eq.~(\ref{u_y}) to the value given. 

The relevance or irrelevance of $H_{2q_F}$ depends on the stiffness of
the stripes measured, roughly, by the dimensionless parameter $Y l^4 /
\tilde{u}_{\rm H}(\pi / \Lambda)$. If it is small enough, then high-$q$
and $\omega$ fluctuations of $\phi^j_\pm$ and $u^j_y$ (either quantum or
thermal) are sufficiently strong to cause effective averaging out of the
cosines in $H_{2q_F}$. As a result, the coefficient $\lambda$
renormalizes to zero in the low $q$ and $\omega$ limit. In contrast, for
rigid stripes, the averaging of the cosine terms is not important. The
perturbative renormalization group (RG)
analysis~\cite{MacDonald_00,Fertig_99,Yi_00,Lopatnikova_01} allows to
make this argument more precise.~\footnote{These four works are in
mutual agreement on how the RG procedure needs to be formulated. A
different type of RG analysis suggested in Ref.~\cite{Barci_01} is
believed to be in error.} However, to decide between the crystal or the
smectic behavior, the RG requires accurate estimates for $Y$ and other
parameters of the edge theory as an input. Attempts to extract such
parameters from the TDHFA for the case of current experimental interest,
$N \sim 2$, led to contradictory statements in the
literature~\cite{MacDonald_00,Yi_00,Lopatnikova_01}. In this regard, we
wish to reiterate that at $N \sim 2$ the stripe phase is very fragile
and faces a strong competition from other quantum Hall states.
Therefore, the Hartree-Fock and its time-dependent extensions are not
quantitatively reliable. At the present stage the controlled theoretical
analysis can be envisioned only for large $N$, building on the work of
Moessner and Chalker~\cite{Moessner_96}. Note that the ratio $Y l^4 /
\tilde{u}_{\rm H}(\pi / \Lambda)$ becomes an $N$-independent number of
the order of unity at $N \gg 1$ and $r_s \sim 1$.

\begin{figure}
\includegraphics[width=1.6in,bb=204 426 429 632]{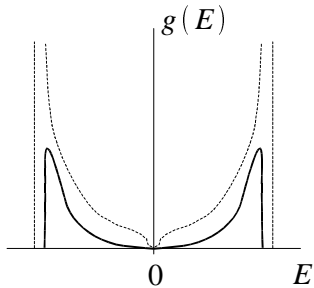}
\caption{
The sketch of the true tunneling density of states (solid line) and
its Hartree-Fock approximation (dashed line).
\label{Fig_DOS}
}
\end{figure}

Concluding this section, let us discuss the tunneling into the stripe
phase from the normal metal. Two setups can be imagined: tunneling from
a bulk metal or from an STM tip. In the first case the electron-electron
interactions are screened by the metal, in the second they remain
long-range. The differential tunneling conductance $G_T = d I / d V$ at
bias voltage $V$ is proportional to the tunneling density of states $g(e
V)$, which is defined by $g(E) = (1 / \pi) \Im{\rm m} \tilde{G}_R(E /
\hbar)$, $\tilde{G}_R(\omega)$ being the single-electron Green's
function,
\begin{equation}
     G_R(t) = -i \theta(t) \langle \Psi(0) \Psi^\dagger(t) \rangle.
\label{G_R}
\end{equation}
Here $\Psi^\dagger$ and $\Psi$ are the electron creation and
annihilation operators. The crude overall behavior of $g(E)$ can be
obtained at the Hartree-Fock level. In the simplest case of the smectic
(unmodulated stripes), $g(E)$ is determined by the slope of the
self-consistent potential $E(X)$ that defines the energies of the
quasiparticle states $|X\rangle$,
\begin{equation}
g_{\rm HF}(E) = |4 \pi l^2 \Lambda (d E / d X)|^{-1}.
\label{g_HF}
\end{equation}
The tunneling density of states calculated according to this formula is
nonvanishing in a finite interval of width $\sim E_{\rm ex}$ around $E =
0$. At both ends of such an interval one finds two divergencies (van Hove
peaks). The interior of the interval can be described as a shallow
pseudogap, see Fig.~\ref{Fig_DOS}. The negative-$E$ peak corresponds to
tunneling into the states with $X$'s in the middle of the filled stripes
and the positive-$E$ one --- into the middle of the empty stripes.
Conversely, at low energies (low $V$) the tunneling can only occur into
points in a vicinity of stripe edges~\cite{Fogler_96}. The Hartree-Fock
results for $g(E)$ are certainly not exact. However, the van Hove peaks
and the pseudogap are expected to be true features and the Hartree-Fock
estimate for the energy separation between the peaks should be quite
reliable. The largest deviations from the Hartree-Fock predictions
should occur at low $E$, which we now address. We will use the evolution
in the imaginary time $\tau = i t$ picture common in quantum tunneling
problems.

%
%

Since the edges are effectively far away from one another, the incoming
electron is initially accommodated into the closest edge, say, $j_+$. It
produces a compact charge disturbance with a high Coulomb energy. The
charge has to gradually spread over the entire area. This takes place by
the emission of the magnetophonons. (In the context of the edge-state
theories they are traditionally called edge magnetoplasmons). What is
missing in the Hartree-Fock picture is precisely this relaxation from a
point-like disturbance into an uniform state with an added quasiparticle
of the given small energy $E$. Its net result is a suppression of the
true $g(E)$ compared to $g_{\rm HF}(E)$. To calculate the suppression
factor, one can use the bosonization prescription $\Psi \propto \exp(i
\phi^j_+)$~\cite{Wen_92}, which entails
\begin{equation}
\tilde{G}_R(i \omega_n) \propto \int\limits_0^\infty
 d \tau \exp\Bigl[i \omega_n \tau
           - \frac12 \langle \phi^j_+(0) \phi^j_+(\tau)
           - \phi^j_+(0) \phi^j_+(0) \rangle\Bigr].
\label{G_R_bosonized}
\end{equation}
This expression can be evaluated with the help of the effective
action~(\ref{L_stripes}) and then analytically continued to real
frequencies $i \omega_n \rightarrow \omega + i 0$~\cite{Mahan_book}. For
the smectic, such a calculation has been performed by Lopatnikova~{\it
et al.\/}~\cite{Lopatnikova_01}. They obtained $I(V) \propto
\exp(-\sqrt{V_0 / V})$ and $I(V) \propto \exp[-\ln^2 (V_0 / V)]$ for the
Coulomb and short-range interaction case, respectively. The last formula
was also obtained by Barci~{\it et al.\/}~\cite{Barci_01}. As one can
see, the tunneling conductance vanishes at zero voltage, which is a
typical result for tunneling into a correlated electron system. For the
crystal, the calculation is more difficult because the action is
essentially nonlinear [the harmonic
approximation~(\ref{H_2q_F_expanded}) is not adequate for the tunneling
problem]. It has not been reported in the literature. On the physical
grounds, we expect the tunneling conductance to be exactly zero unless
$e V$ exceeds the energy needed to nucleate a phase slip $u_y
\rightarrow u_y + a$ (an analog of an interstitial in conventional
crystals).

\section{Pinning of stripes by disorder}
\label{Pinning}

Disorder in the form of randomly positioned impurities is expected to
strongly affect the degree of the positional, orientational, and
topological orders in the stripe phase at low $T$. According to modern
understanding, none of these survives in the thermodynamic limit in two
dimensions. However, in the most interesting case of weak disorder, the
details of how each type of order is destroyed are subtly different. Let
us start the discussion with the case of a smectic. Here the positional
order is limited primarily by random elastic distortions. They can be
characterized by the roughness function
\begin{equation}
f(r) = \langle [u({\bf r}^\prime) - u({\bf r}^\prime + {\bf r})]^2
\rangle,
\label{f}
\end{equation}
so that the positional correlation length $\xi_P$ can be defined by the
condition $f(\xi_P) = \Lambda^2$. On the other hand, the correlation
lengths for the topological order $\xi_T$ and orientational order
$\xi_O$ are set by the mean distance between the disorder-generated
dislocations and disclinations, respectively. Note that an appreciable
magnetotransport anisotropy can be detected only if the sample size is
smaller than $\xi_O$. This explains why samples of macroscopic
dimensions $L \sim 0.25\,{\rm mm}$ show no anisotropy unless they are
extremely pure~\cite{Lilly_99,Du_99}. In addition to weakness of
disorder, the gigantic $\xi_O$ in these experiments may have originated
from a small bare anisotropy, which favored the alignment of the stripes
along the $[1 1 0]$ direction. Indeed, an external
anisotropy qualitatively modifies the smectic state properties by
generating a finite tilt modulus $Y_\perp$. Correspondingly, the elastic
part $H_{el}$ of the Hamiltonian becomes
\begin{equation}
 H_{el} = \frac{Y}{2} \Bigl[\partial_x u + \frac12 (\nabla u)^2\Bigr]^2
 + \frac{Y_\perp}{2} (\partial_y u)^2
 + \frac{K}{2} (\partial_y^2 u)^2.
\label{H_smectic_anisotropy}
\end{equation}
The full Hamiltonian $H = H_{el} + H_{pin}$ also includes the
interaction with disorder. It can be approximated by
$H_{pin} = \int \exp [i q_* u({\bf r})] V({\bf r})$ where $V({\bf r})$
is a random short-range potential. The behavior of the static
correlation functions, such as $f(r)$, as functions of $r$ and disorder
strength has been evaluated within this model by Scheidl and
von~Oppen~\cite{Scheidl_01} building on the theory developed earlier for
other pinned systems~\cite{Villain_84,Radzihovsky_99}. Scheidl and
von~Oppen suggested that in the current experiments the characteristic
lengthscales satisfy the chain of inequalities $\xi_P < \xi_T < L <
\xi_O$, so that the stripe phase resembles an instantaneous snapshot of
a nematic with no long-range positional order and a number of frozen-in
dislocations.

Besides decorrellating the stripe positions, another interesting
disorder effect in the smectic phase can be inducing quenched random $2
q_F$-modulations along the stripes, as suggested by Yi~{\it et
al.\/}~\cite{Yi_00}. This is similar to the effect of impurities on
a 1D ``Wigner crystal''~\cite{Glazman_92} (where without impurities the
density is uniform because of strong quantum fluctuation). To properly
describe this effect and its consequences, a comprehensive {\it
quantum\/} theory of pinned smectics is required.

The pinning effects in the stripe crystal phase are qualitatively
similar because $H_{el}$ is already not too different from the elastic
Hamiltonian of an anisotropic crystal. The main difference is the
doubling of the number of components of the elastic displacement field
${\bf u} = \{u_x, u_y\}$.

The combined effects of finite temperature and disorder have not been
investigated in any detail, but based on the extensive work done in
other contexts~\cite{Blatter_94,Nattermann_00} we may expect a thermal
depinning at a certain ``glass transition temperature.''

Exceedingly interesting topic is the dynamics of pinned electron
(liquid) crystals in high magnetic field. First of all, there are number
of open questions in the general theory of pinned manifolds (such as
CDWs, vortex lattices, magnetic domains, {\it etc.\/}). The current
interests in this subject area include dislocation dynamics and
non-equilibrium states (e.g., high electric field sliding phases). In
addition to that, there are interesting dynamical effects unique to
quantum Hall systems. For instance, in the recent years it was
experimentally discovered~\cite{Li_97} that even the linear response of
a pinned Wigner crystal in a magnetic field is dramatically different
from conventional expectations~\cite{Fukuyama_78}. Typically, one
anticipates the response to be dominated by low-frequency magnetophonon
modes, which in the presence of pinning become localized wavepackets.
Since the disorder is different in different parts of the sample, a
considerable inhomogeneous broadening is expected. Instead, a sharp
resonance-like mode in the microwave absorption spectrum appears
(typically around $1\,{\rm GHz}$). Assuming that the quasiclassical
description is adequate, Huse and the present
author~\cite{Fogler_Huse_00} explained the appearance of a narrow mode
as a combined effect of the special Lorentz-force dynamics and
long-range Coulomb interaction. However, the theory was done for the $T
= 0$ while the experimental line width seem to be dominated by thermal
broadening. In fact, the thermal broadening is much more pronounced than
one would naively expect and may reflect some nontrivial physics
overlooked in a formulation based on harmonic elasticity
theory~\cite{Fogler_Huse_00}. The similar pinning resonance should exist
in a stripe crystal as well and may even persists in a smectic phase.
Since at high LLs all the energy scales are smaller, the pinning mode
should have a lower frequency, perhaps, of the order of 100 MHz at $\nu
= 9 / 2$ for the kind of samples used in Refs.~\cite{Lilly_99,Du_99}.

\section{Magnetotransport in the stripe phase}
\label{Transport}

The magnetotransport in a partially filled Landau level is a problem of
two extremes. In a perfectly clean system, it is completely trivial and
amounts to the uniform sliding motion of all the particles in the
direction perpendicular to the applied electric field, leading to
$\sigma_{x x} = 0$, $\sigma_{x y} = (e^2 / h) \nu$. On the other hand,
in the presence of disorder, the magnetotransport immediately becomes
very complicated. Especially for the stripe phase, the magnetotransport
theory is very incomplete at the moment. For instance, the transport in
the nematic phase is a total enigma. Certain progress has
been achieved regarding the crystal and the smectic phases. As discussed
above, pinning by random impurities eliminates the possibility of a
global sliding motion of the electron liquid under the action of an
external electric field. As a result, the stripe crystal is an
insulator. In the smectic, gapless edge states may exist in which case
the edge transport is possible. The quasiparticles travel along the
edges in the direction prescribed by the edge chiralities and also be
scattered between the edges by impurities. The scattering probability
depends on the distance between the edges. For a general filling
fraction $\nu_N$ the widths of the filled (electron) and empty (hole)
stripes are unequal, and so the corresponding mean-free paths are also
different, $l_e \neq l_h$. On a large scale the quasiparticle motion
resembles a random walk with unequal elementary steps: along the
stripes, it is a properly weighted average of $l_e$ and $l_h$, across
the stripes it is the interstripe separation $\Lambda$. If we assume
that (a) the stripes are continuous and are all oriented along the
$\hat{\bf y}$-direction, (b) the scattering occurs only between nearest
edges, and (c) quantum localization effects can be neglected, then a
simple calculation leads to the following formulas for the components of
the conductivity tensor~\cite{MacDonald_00,Oppen_00b}:
\begin{equation}
 \sigma_{x x}=\frac{e^2}{h} \frac{\Lambda^{-1}}{l_e^{-1} + l_h^{-1}},
\quad
 \sigma_{y y}=\frac{e^2}{h} \frac{\Lambda}{l_e + l_h},
\quad
 \sigma_{x y} = \frac{e^2}{h} \Bigl(N + \frac{l_e}{l_e + l_h}\Bigr).
\label{sigma}
\end{equation}
At the symmetric point $l_e = l_h$, the transport anisotropy ratios are
given by $R = \sigma_{y y} / \sigma_{x x} = \rho_{x x} / \rho_{y y} =
l_e^2 / \Lambda^2$, which is large for dilute impurities. However, there
are reasons to doubt that Eq.~(\ref{sigma}) directly applies to the
experimental practice. Indeed, from the empirical estimate $R
\sim 7$ one would conclude that $l_e \sim 2.6 \Lambda \sim 0.2\,\mu{\rm
m}$, which seems too short for the extreme purity samples used. The most
probable reason for reduction of the anisotropy ratio compared to the
idealized formula~(\ref{sigma}), is the absence of high degree of the
orientational and topological orders. As explained above, the pinning by
disorder introduces pronounced elastic deformations as well as
dislocations and disclinations into the perfect stripe pattern. Thus, in
a more realistic model the above condition (a) should be abandoned. This
leads to a view of the low-temperature stripe phase is a collection of
elongated puddles of $\nu_N = 1$ state preferentially aligned in a
certain direction (a kind of frozen nematic). The conductivity tensor
can no longer be calculated in any simple way; however, it satisfies the
generalized Dykhne-Ruzin semicircle law~\cite{Dykhne_94}
\begin{equation}
 \sigma_{x x} \sigma_{y y} + (\sigma_{x y} - \sigma_0)^2 = (e^2 / 2 h)^2,
\quad \sigma_0 \equiv (e^2 / 2 h)(2 N + 1),
\label{semicircle}
\end{equation}
pointed out by von~Oppen, Halperin, and Stern~\cite{Oppen_00b}.
In addition, exactly at half-filling $\sigma_{x y} = \sigma_0$ and
the product rule~\cite{Dykhne_94,MacDonald_00,Oppen_00} applies:
\begin{equation}
 \sigma_{x x} \sigma_{y y} = (e^2 / 2 h)^2.
\label{product_rule}
\end{equation}
Both relations agree quite well with the low-temperature experimental
data at $\nu = 9/2$~\cite{Eisenstein_01}, which is encouraging. However,
one has to be careful with drawing conclusions from such a comparison.
The status of Eqs.~(\ref{semicircle}) and (\ref{product_rule}) in the
quantum domain is poorly understood and is even somewhat compromised by
the lack of agreement with the data from dirtier samples. The situation
is complicated by a unconventional behavior of the resistivity peak
at $\nu = 9 / 2$ (low-$T$ width saturation), precisely where the product
rule is supposed to apply.

At higher temperatures where the CDW amplitude is small, the stripe
edges are not well defined objects. It is better to describe the system
as a ``compressible liquid.'' A new type of transport theory is needed
in this regime. It is reasonable to assume, for example, that the
stripes can be depinned already by a vanishingly small electric field,
so that a global sliding of the electron liquid becomes possible. If
disorder is sufficiently smooth, it would be similar to a flow of a
viscous 2D liquid across a disordered substrate~\cite{Fogler_unpub}.

Sufficiently close to $T_c$ the dislocations must become highly mobile
so that the stripe phase should behave as a nematic on all relevant
lenghtscales. A phenomenological approach to the magnetotransport in
this regime was advocated by Fradkin {\it et al.\/}~\cite{Fradkin_00}.
They argue on the basis of symmetry that the deviation of the
resistivity ratio $R = \rho_{x x} / \rho_{y y}$ from unity is
proportional to the degree of the orientational order in the nematic.
Under a simplifying assumption that the Frank constants are equal, they
map the problem onto a 2D $X$-$Y$ model in the presence of a small
orienting field. Using the Monte-Carlo results for the latter model,
they fitted the experimental data for $R$ with the help of
two free parameters.

\section{Other topics}
\label{Other}

The space constraints force us to switch to a telegraphic-style
overview of the remaining topics explored in connection
with the CDW phases in high LLs.
 
One of those is the bare anisotropy that aligns the stripes in the $[1 1
0]$ direction. Its origin remains unclear. A careful study by
Cooper~{\it et al.\/}~\cite{Cooper_01} has ruled out a few earlier
suggestions that included substrate morphology~\cite{Willett_xxx},
vicinal steps from a slight miscut of the wafer, and a small anisotropy
of the effective mass~\cite{Takhtamirov_xxx,Rosenow_01}. One of the
remaining viable alternatives is piezoelectric effects~\cite{Fil_00}.
The orienting effects of the in-plane magnetic field was further studied
in wide quantum wells~\cite{Pan_wide_well_00}. The combination of the
empirical findings and the Hartree-Fock calculations suggest that the
anisotropy is indeed tiny, $\sim 1\,{\rm m K}$ per electron.
The corresponding analysis for the $p$-type GaAs~\cite{Shayegan_00} has
also been performed~\cite{Kim_00}. It is more involved because of the
band-structure anisotropy. The preferred stripe orientation was
predicted to sensitively depend on various details such as the width of
the quantum well.

The time-dependent Hartree-Fock calculations of the collective modes of
the stripe crystal has been done by Fertig~{\it et
al.\/}~\cite{Cote_00,Yi_00}. At low $q$'s the results are in
agreement with the general structure predicted by the effective
theories, Secs.~\ref{Liquid_crystals} and \ref{Edge_models}. At higher
$q$ a new feature is revealed: a roton-like minimum at $q \approx q_*$,
which may be thought of as a precursor to the stripe-bubble transition. 

Stripe phases in bilayer systems have been studied by Brey and
Fertig~\cite{Brey_00} and also by Demler~{\it et
al.\/}~\cite{Demler_xxx}. The additional layer degree of freedom leads
to a rich phase diagram. Some of that physics may be operational in wide
quantum wells~\cite{Pan_wide_well_00}.

The emergence of the CDW indicates some sort of instability of the
conventional uniform FQHE states at high LLs. Such an instability was
indeed found in numerical studies of Scarola~{\it et
al.\/}~\cite{Scarola_01} who relied on the composite-fermion theory. It
was also discussed within the phenomenological mean-field
composite-boson formulation~\cite{Rosenow_01}.

\section{Conclusions}
\label{Conclusions}

The physics of high LLs has fledged into a fast growing and vibrant
field. Despite a significant progress, a multitude of open problems
remains. Some of them are brand new (the existence of qualitatively
novel states of matter such as quantum nematic), others are venerable
quantum Hall problems with a new twist (the integral quantum Hall
transition in the stripe phase). The real progress in understanding
these intriguing yet difficult problems is likely to be tied with future
experimental advances. It should be mentioned that the magnetotransport
studies have certainly not exhausted their potential. Ongoing fantastic
achievements in sample fabrication continue to deliver the ever more
perfect systems, which should expedite the discovery of new phases. A
fresh example is the aforementioned group of new insulating states in
the $N = 1$ LL of $3.1 \times 10^7\,{\rm c m}^2 /{\rm V s}$ mobility
sample~\cite{Eisenstein_xxx} (Sec.~\ref{Experimental_evidence}). Besides
relying on increasing sample quality, it may be interesting to
experiment with different sizes of already available samples in search
for the sequence of ``spaghetti'' phases sketched in
Fig.~\ref{Fig_spaghetti}. On the other hand, finite-frequency tools,
such as microwaves, surface acoustic waves, and inelastic light
scattering will provide other invaluable information inaccessible by the
dc magnetotransport. Finally, the real-space
imaging~\cite{Finkelstein_00,Yacoby_00} of the stripes and bubbles would
be the most definitive proof of their existence.

It is a safe bet that we will see a great number of surprises and new
developments in this area.

\section{Acknowledgements}

This work is supported by the MIT Pappalardo Fellowships Program in
Physics. I would like to thank A.~A.~Koulakov, B.~I.~Shklovskii, and
V.~M.~Vinokur for previous collaboration during which the key ideas and
insights surveyed in these notes have emerged. I also wish to thank
B.~I.~Shklovskii for valuable comments on the manuscript.

\end{document}